\documentclass[12pt]{article} 
\usepackage[a4paper, margin=1in]{geometry}  % Set 1-inch margins
\usepackage{setspace}  % For double spacing
\usepackage{lmodern}  % Improved font rendering
\usepackage{hyperref}  % Hyperlinks
\usepackage[numbers,authoryear]{natbib}  % For author-year citations
\usepackage{amsmath, amsfonts, amssymb}
\usepackage{booktabs}
\usepackage{graphicx}
\usepackage{multicol, multirow}
\usepackage{pdflscape}
\usepackage{caption}
\usepackage{float}
\title{\textbf{Inverse sampling intensity weighting for preferential sampling adjustment}}
    \author{Thomas W. Hsiao\textsuperscript{1}\thanks{Corresponding author: Thomas W. Hsiao, email: thsiao3@emory.edu}
 \and Lance A. Waller\textsuperscript{1}
}
\date{
    \textsuperscript{1}Department of Biostatistics and Bioinformatics, \\ 
    Rollins School of Public Health, Emory University \\
    Atlanta, Georgia, USA \\ \vspace{4mm}
    \today  % Automatically inserts today's date
}

\begin{document}
\begin{singlespace}
\maketitle
\thispagestyle{empty} % No page number on the title page

\begin{abstract}
Traditional geostatistical methods assume independence between observation
locations and the spatial process of interest. Violations of this independence assumption are referred to as preferential sampling (PS). Standard methods to address PS rely on estimating complex shared latent variable models and can be difficult to apply in practice.
We study the use of inverse sampling intensity weighting (ISIW) for PS adjustment in model-based geostatistics. ISIW is a two-stage approach wherein we estimate the sampling intensity of the observation locations then define intensity-based weights within a weighted likelihood adjustment. Prediction follows by substituting the adjusted parameter estimates within kriging. We introduce an implementation of ISIW based on the Vecchia approximation, enabling computational gains while maintaining strong predictive accuracy. Interestingly, we found that ISIW outpredicts standard PS methods under misspecification of the sampling design, and that accurate parameter estimation had little correlation with predictive performance, raising questions about the conditions driving optimal implementation of kriging-based predictors under PS. Our work highlights the potential of ISIW to adjust for PS in an intuitive, fast, and effective manner. We illustrate these ideas on spatial prediction of lead concentrations measured through moss biomonitoring data in Galicia, Spain, and $\text{PM}_{2.5}$ concentrations from the U.S. EPA Air Quality System network in California. 

\end{abstract}

\vspace{1em}
\noindent \textbf{Keywords:} inverse sampling intensity weighting, ISIW, model-based geostatistics, preferential sampling, kriging, point process
\end{singlespace}
\newpage

\setcounter{page}{1}
\pagenumbering{arabic}
\section{Introduction}
%\linenumbers

The field of geostatistics comprises a set of methods to infer properties and predict unknown values (or functions) of a spatially continuous process $\{S(\textbf{x}):\textbf{x}\in\mathcal{D}\subseteq \mathbb{R}^2\}$ from a discrete set of observation locations, denoted $\textbf{X} = \{\textbf{x}_1, \textbf{x}_2, \ldots, \textbf{x}_n\}$. Model-based approaches estimated via maximum likelihood estimation (MLE) typically assume $S$ is a realization from a random stochastic process, and subsequent inference and prediction treat $\textbf{X}$ as fixed \citep{diggle_model-based_1998}. This assumption further implies that the sampled locations are independent of the spatial process of interest, or in symbols, $[\textbf{X},S]=[\textbf{X}][S]$, where square brackets indicate the probability distribution. Violations of this independence assumption are referred to as \textit{preferential sampling} (PS). Several studies have shown that neglecting PS can introduce substantial bias in geostatistical inference and prediction \citep{diggle_geostatistical_2010, pati_bayesian_2011, gelfand_effect_2012}. 

Early works in PS examined $\textbf{X}$ under the lens of a marked point process. \cite{schlather_detecting_2004} designed two Monte Carlo tests to detect dependence between marks and their locations. \cite{ho_modelling_2008} introduced the classic PS adjustment model, referred to here as the \textit{shared latent process} (SLP) model, and derived several of its properties. Full estimation and prediction procedures for the SLP model were formally established in the seminal work by \cite{diggle_geostatistical_2010}. 

The SLP framework induces dependence between the spatial process of interest and the sampling locations through a shared variable approach.  Let $\mu, \alpha, \beta,$ and  $\tau^2$ be scalars in $\mathbb{R}$, $\textbf{Y}$ denote the $n\times 1$ random vector of observed values sampled at locations $\textbf{X}$, $S$ the shared latent process, $S(\textbf{X})$ the values of $S$ evaluated at locations $\textbf{X}$, $IPP$ an inhomogeneous Poisson process, $GP$ a Gaussian process, $C_\theta(\cdot,\cdot)$ a stationary covariance function indexed by parameter vector $\theta$, and $\lambda(\cdot)$ the intensity function of $\textbf{X}$ conditional on $S$. Then the SLP model is as follows:

\begin{equation} \label{eqn:slp}
\begin{split}
    [\textbf{Y}|\textbf{X}, S] \sim N_n(\mu \textbf{1}_n + S(\textbf{X}), \tau^2I_n), \\
    [\textbf{X}|S] \sim IPP(\lambda), \\
    \lambda(\textbf{x})=\exp(\alpha+\beta S(\textbf{x})), \\
    S \sim GP(0, C_\theta(\cdot, \cdot)).\\
\end{split}
\end{equation}

While the SLP framework has been extended to handle covariates in the mean model, non-Gaussian likelihoods, and multiple shared latent processes \citep{pati_bayesian_2011, watson_general_2019}, for the purposes of illustration, we focus on the original formulation of a constant mean, Gaussian likelihood, and one shared latent process. Under the SLP model, sampling locations are no longer assumed to be fixed and are instead a realization from a log Gaussian Cox process (LGCP). The latent process $S$ is assumed to follow a zero mean second order stationary Gaussian process (GP). The SLP takes its name from the presence of $S$ in both the mean model of $\textbf{Y}$ and the intensity function of $\textbf{X}$. The parameter $\beta$ can be interpreted as a single PS coefficient controlling the preferentiality of the SLP. The sign of $\beta$ determines the direction of the preference, while its magnitude determines the likelihood of sampling extreme values. 

Subsequent research on PS methods has remained largely faithful to the SLP model. \cite{pati_bayesian_2011} extended the SLP model structure to a flexible Bayesian framework estimated by MCMC, and proved the posterior consistency of each of the mean, covariance and PS parameters under increasing domain asymptotics. \cite{gelfand_effect_2012} also used Bayesian estimation, and introduced a framework to compare prediction surfaces under PS between different methods. A surprising finding from their simulation analysis was that inclusion of an informative covariate was not sufficient to correct for predictive bias. This empirical finding highlights the importance of the SLP model even when informative covariates explaining the dependence between $\textbf{X}$ and $S$ are available. \cite{dinsdale_methods_2019} significantly improved the computational efficiency of the SLP model relative to the previous simulation-based methods by means of the stochastic partial differential equations (SPDE) approach \citep{lindgren_explicit_2011} combined with a Laplace approximation implemented in the Template Model Builder (TMB) \texttt{R} package \citep{kristensen_tmb_2016}. \cite{watson_general_2019} also fitted the SLP model using the SPDE approach but with an integrated nested Laplace approximation implemented in the \texttt{R-INLA} software \citep{rue_approximate_2009}, which enjoys comparable computational gains to the approach in \cite{dinsdale_methods_2019}. The authors further defined a framework to model PS spatio-temporal data and better emulate the evolution of $\textbf{X}$ over time compared to the original SLP model defined in \eqref{eqn:slp} above. 

Several other contributions extend the SLP framework to explore a rich set of important applications, including air pollution monitoring \citep{lee_constructing_2011, lee_impact_2015}, species distribution modeling \citep{manceur_inferring_2014, fithian_bias_2015, conn_confronting_2017, pennino_accounting_2019, gelfand_preferential_2019, fandos_dynamic_2021}, disease surveillance \citep{rinaldi_sheep_2015, cecconi_preferential_2016, conroy_shared_2023}, phylodynamic inference \citep{karcher_quantifying_2016}, hedonic modelling \citep{paci_spatial_2020}, bivariate spatial data \citep{shirota_preferential_2022}, spatially-varying PS \citep{amaral_model-based_2023}, optimal design under PS \citep{ferreira_optimal_2015}, space-filling designs \citep{ferreira_geostatistics_2020, gray_design_2023}, a hypothesis test to detect PS in spatio-temporal data \citep{watson_perceptron_2021}, and exact Bayesian inference for the SLP model \citep{moreira_analysis_2022, moreira_presence-only_2024}. The SLP's flexibility in addressing sampling bias has driven the popularity of model-based approaches that account for the inherent dependence between spatial processes and their observation locations. 

Despite these advantages, the SLP framework contains significant limitations that complicate its application. Few software packages and out-of-the-box implementations exist for SLP estimation, often coming in the form of custom INLA or MCMC sampler code. Furthermore, the SLP model is challenging to integrate with modern spatial Gaussian process approximation techniques, many of which have become the standard in spatial data analysis due to their scalability and feasibility, offering computationally efficient alternatives to the MLE while maintaining high accuracy \citep{heaton_case_2019}. Among these methods, only the SPDE approach has a well-documented implementation of a PS solution, and incorporating the SLP model into any other method would take a substantial effort. The SLP model requires the nontrivial task of jointly modeling both the response and observation likelihood. Prediction based on the SLP model can also suffer from high computational cost. While parameter estimation for the SLP model may scale well with the SPDE approach, prediction depends on summarizing the posterior distribution of $[S|\textbf{X}, \textbf{Y}]$ for each prediction point, which can be infeasible even on a moderately sized grid for both the Laplace approximation and MCMC. 

 An alternative strategy to the SLP is to incorporate dependence through the sampling intensities $\lambda$ of the observations, rather than the entire likelihood of $\textbf{X}$. The two main potential uses of sampling intensities are as a covariate in the mean model of $\textbf{Y}$ or as inverse weights in a likelihood adjustment \citep{vedensky_look_2023}. It remains unclear how well these methods work in geostatistical applications. A potential advantage of sampling intensity methods is their robustness to the form of PS relative to the SLP framework, which requires full parametric specification of the likelihood of $\mathbf{X}$. 

The biggest impediment to the use of the sampling intensity is the need to estimate this intensity from the observed locations. Unlike in survey methodology where survey weights are fixed and known, sampling weights in geostatistical applications are largely unknown to the investigator and must be estimated. Unfortunately, nonparametric kernel smoothing estimators of the intensity  \citep{diggle_kernel_1985, berman_estimating_1989} have few theoretical guarantees and are not consistent for the true intensity without knowledge of spatial covariates \citep{guan_consistent_2008}. 

Even so, there is preliminary evidence that methods using nonparametrically estimated sampling intensities can still mitigate the effects of PS. \cite{mateu_accounting_2012} noted how the Bayesian SLP model parameters could be estimated without MCMC by replacing the shared latent process term in the mean model with some known function of the sampling intensity $g(\lambda)$. In another example, \cite{zidek_reducing_2014} provide an approach using estimated weights for air pollution monitoring. Instead of working in a continuous study region, however, these authors considered a finite superpopulation of possible sampling sites and the probability of selecting any site over time was modeled by a logistic regression. The estimated inverse probabilities were then used as weights in a Horvitz-Thompson style design-based estimator for unbiased estimation of parameters under PS. \cite{schliep_correcting_2023} also considered a finite superpopulation and similarly deviated from traditional model-based geostatistics by using estimated sampling intensity weights to recover the estimated parameter values and kriging variance had the model been fit on the superpopulation, rather than estimate the parameters of the spatial process $S$. Finally, \cite{vedensky_look_2023} conducted a simulation experiment comparing a univariate marginal composite likelihood (CL) weighted by inverse sampling intensities estimated by the \texttt{MASS::kde2d} function in \texttt{R} to the SLP and unweighted CL, and showed improved performance compared to no adjustment. We will refer to methods using inverse sampling intensities as weights in a weighted likelihood adjustment as \textit{inverse sampling intensity weighting} (ISIW). 

Despite preliminary investigations into ISIW for PS adjustment, specifics regarding the effectiveness of such approaches within model-based geostatistics remain largely unknown. In particular, it is unclear whether they reliably estimate both mean \textit{and} covariance parameters for a latent spatial process on a 2D continuous surface. Composite likelihoods beyond pairwise difference and univariate marginal have also not been explored for ISIW. Finally, the robustness of ISIW and the SLP to misspecification has not been well-studied. In the sections below, we provide an expanded evaluation of ISIW methods, comparing the performance of MLE, SLP, and ISIW across multiple random fields and PS designs.

Our work proceeds as follows. We start off in Section \ref{sec:aim2_background} by providing background on model-based geostatistics and introduce the key methods we use as a basis for ISIW. In Section \ref{sec:aim2_ISIW}, we present the implementation of ISIW in detail. In Section \ref{sec:aim2_simulation}, we conduct a comparison analysis of the MLE, SLP, and ISIW through a set of simulation studies. In Section \ref{sec:aim2_galicia} we apply the same methods to the famous Galicia moss dataset which has been the dataset of choice when investigating PS, as well as a dataset of air pollution measurements taken from the Air Quality System (AQS) in California. In Section \ref{sec:aim2_discussion} we finish with final remarks and discuss the implications of our work, and outline future directions for continuing investigation. 

\section{Model-based geostatistics}\label{sec:aim2_background}
\subsection{Maximum likelihood estimation}
Under non-preferential sampling (NPS), we define a model for the Gaussian observations $\textbf{Y}$ and underlying Gaussian process $S$ following the SLP framework in \eqref{eqn:slp}, but we drop the point process likelihood for $\textbf{X}$. The standard geostatistical model under NPS follows %Let $\mu$ and $\tau^2$ be the mean and nugget parameters in $\mathbb{R}$, $\textbf{Y}$ denote the $n\times 1$ random vector of observed values sampled at \textit{fixed} locations $\textbf{X}$, $S$ the latent Gaussian process, $S(\textbf{X})$ the values of $S$ evaluated at fixed locations $\textbf{X}$, and $C_\theta$ a covariance function indexed by parameter vector $\theta$. Then our hierarchical model for the observations follows
\begin{equation*} \label{eqn:aim2_MLE}
\begin{split}
    [\textbf{Y}|\textbf{X}, S] \sim N_n(\mu \textbf{1}_n + S(\textbf{X}), \tau^2I_n), \\
    S \sim GP(0, C_{\boldsymbol{\theta}}(\cdot, \cdot)).\\
\end{split}
\end{equation*}
We further assume the covariance function $C_\theta$ follows a stationary isotropic Mat\'ern covariance defined as
\begin{equation*}\label{eqn:aim2_matern}
    C_\theta(h) = \sigma^2 \frac{2^{1-\nu}}{\Gamma(\nu)}\left(\sqrt{2\nu}\frac{h}{\phi}\right)^\nu K_\nu\left(\sqrt{2\nu} \frac{h}{\phi}\right),
\end{equation*}
where $\theta:=(\sigma^2, \nu, \phi)$ defines the covariance parameter vector containing the variance, smoothness, and range parameters respectively, $h$ denotes the Euclidean distance between two points of observation, and $K_\nu$ is the modified Bessel function of the second kind. By convention, the smoothness parameter $\nu$ is assumed to be known and fixed {\em a priori} before estimation. Therefore, the parameters of interest are $\boldsymbol\psi:=(\mu, \sigma^2,\phi,\tau^2)$.  

Because any finite collection of random variables corresponding to point observations of $S$ follows a multivariate Gaussian distribution and \textbf{X} is treated as fixed, the observed data likelihood is then defined by
\begin{equation*} \label{eqn:aim2_MLE_likelihood}
\begin{split}
    [\textbf{Y}, \textbf{X}] \propto N_n(\mu \textbf{1}_n, \Sigma_n(\theta) + \tau^2I_n), \\
\end{split}
\end{equation*}
where $\Sigma_n$ is the covariance matrix with $ij$th entry equal to $C_\theta(\lVert\textbf{x}_i- \textbf{x}_j\rVert)$. Estimation proceeds by optimization of the likelihood with respect to the parameter vector $\boldsymbol{\psi}$. 

The main drawback for practical use of the MLE is its computational burden. Evaluating the likelihood requires an inversion of the covariance matrix, which has $O(n^{3})$ time and $O(n^2)$ space complexity making it infeasible for applications for moderately sized $n$. Recent advancement in modern spatial statistics has focused on GP parameter estimation using approximations which massively reduce computational burden in exchange for marginal decreases in efficiency. In the next section, we discuss the Vecchia approximation, one such alternative likelihood approach that achieves high computational and statistical efficiency. 

\subsection{Vecchia approximation}
The Vecchia approximation is a specific case of composite likelihood \citep{varin_overview_2011} based on the observation that the joint distribution of a random vector can be decomposed as the product of conditional distributions. Let $\textbf{Y}=(Y_1,...,Y_n)^\top \in \mathbb{R}^n$ and $p:\{1,...,n\}\to\{1,...,n\}$ be a permutation mapping which reorders $\textbf{Y}$. Here $f(\cdot; \boldsymbol{\psi})$ denotes the probability density of the corresponding subvector of $\mathbf{Y}$ under parameter vector $\boldsymbol{\psi}$. We define the history of variable $Y_j$ as a random subvector $\textbf{Y}_{h(j)}$ where $h(j)=\{l\in\{1,...,n\}: p(l)<p(j)\}$.  The joint density of $\textbf{Y}$ evaluated at a fixed value $\mathbf{y}$ can then be refactored as
\begin{equation*}
    f(\textbf{y};\boldsymbol{\psi})=f(y_{p(1)};\boldsymbol{\psi}) \prod_{i=2}^n f(y_{p(i)}|\textbf{y}_{h(i)}; \boldsymbol{\psi}).
\end{equation*}

Vecchia observed that much of the information in the conditioning sets with higher indices was likely to be redundant. One could attain a good tradeoff of efficiency for computational gain by decreasing the size of each conditioning set and being judicious about which variables to include. The density would then be approximated as 

\begin{equation}\label{eqn:aim2_vecchia}
    f(\textbf{y};\boldsymbol{\psi}) \approx {f}_V(\textbf{y};\boldsymbol{\psi})= f(y_{p(1)};\boldsymbol{\psi})\prod_{i=2}^n f(y_{p(i)}|\textbf{y}_{q(i)};\boldsymbol{\psi}),
\end{equation}
where $q(i) \subseteq \{p(1),p(2),...,p(i-1)\}$ is the set of indices constituting the conditioning set of $y_{p(i)}$. The Vecchia estimate $\hat{\boldsymbol{\psi}}_V$ maximizes \eqref{eqn:aim2_vecchia}. The Vecchia approximation can also be written in a weighted CL form,
\begin{equation} \label{eqn:aim2_vecchia_CL}
\begin{split}
    \log \mathcal{L}_V(\boldsymbol{\psi};\textbf{y})&=w_1\log f(y_{p(1)};\boldsymbol{\psi})\\
    &+\sum_{i=2}^n w_{1i}\log f(y_{p(i)},\textbf{y}_{q(i)};\boldsymbol{\psi}) \\ 
    &- \sum_{i=2}^n w_{2i}\log f(\textbf{y}_{q(i)};\boldsymbol{\psi}).
\end{split}
\end{equation}

While simple in concept, the Vecchia approximation requires careful selection of three key hyperparameters: 1) the size of the conditioning set, 2) which variables to include in each conditioning set, and 3) the ordering of the variables as determined by $p$. We denote the maximum size of any $q(i)$ as $m$. In all later sections, we choose $m=20$ based on a clear case of diminishing returns in inference and prediction for $m> 20$ observed empirically in \cite{datta_hierarchical_2016}. Default settings in two implementations of the Vecchia approximation (\texttt{GPvecchia} and \texttt{GpGp} \texttt{R} packages) are $m=20$ and $m=30$, respectively \citep{katzfuss_general_2021, guinness_gaussian_2021}. For the choice of which variables to include in the neighborhood $q(i)$, we follow the recommendation given in \cite{vecchia_estimation_1988} by choosing the nearest neighbors to $y_{p(i)}$ measured by Euclidean distance. Ordering for the Vecchia approximation requires additional specification in 2D settings due to the lack of natural ordering in observations of a multidimensional point process. We use the maxmin ordering scheme, which picks a location $p(i)$ sequentially by maximizing the distance to the nearest point in $\{y_{p(1)},y_{p(2)},...,y_{p(i-1)}\}$ and has been shown to approximate the true distribution better than other ordering methods \citep{guinness_permutation_2018, katzfuss_general_2021}.

The Vecchia approximation strikes a favorable balance between computational efficiency and statistical accuracy. Compared with other composite likelihood approaches, it accommodates higher-order dependence, scales well to large datasets, and corresponds to a valid joint density. This is the main reason for considering its use in ISIW methods compared to univariate and pairwise composite likelihoods from previous studies.

\subsection{Kriging}

For many spatial analyses, the main goal is to predict values at unobserved locations. A standard spatial prediction tool is \textit{kriging}, or Gaussian process regression, a function approximation technique with point and variance estimate defined as 

\begin{equation}\label{eqn:aim2_krigall}
    \begin{aligned}
        \hat{S}(\textbf{X}_0) &= \mu + C_\theta(\textbf{X}_0, \textbf{X}_n)^\top\boldsymbol\Sigma_n(\theta)^{-1}(\textbf{Y}-\mu\textbf{1}_n), \\
        \operatorname{Var}\big\{\hat{S}(\textbf{X}_0)\big\} &= C_\theta(\textbf{X}_0, \textbf{X}_0) - C_\theta(\textbf{X}_0, \textbf{X}_n)^\top\boldsymbol\Sigma_n^{-1}(\theta) C_\theta(\textbf{X}_0, \textbf{X}_n),
    \end{aligned}
\end{equation}
where $\textbf{X}_0$ represents unobserved locations to be predicted. In practice, since the parameters are unknown, all parameters in \eqref{eqn:aim2_krigall} are replaced with estimates obtained from any of the aforementioned estimation methods.

It can be shown under the assumption of NPS that kriging is the best linear unbiased predictor (BLUP) \citep{cressie_statistics_2015}. However, this optimality no longer holds under preferential sampling, where even the true values of $\boldsymbol\psi$ used in the kriging equations yield suboptimal predictions \citep{dinsdale_methods_2019}. This is because the optimal predictor is the conditional expectation $\mathbb{E}\big\{S(\mathbf{X}_0)\mid \mathbf{Y}, \mathbf{X}\big\}$, whose distribution is no longer Gaussian as in the fixed case. Nevertheless, kriging can still be used to generate predictions under these non-standard conditions. One potential approach is to solve for the kriging weights under the model defined in \eqref{eqn:slp}. However this optimization does not admit a closed-form expression. We therefore turn to ISIW as a practical approach for finding GP mean and covariance parameters that yield an approximation to the optimal linear kriging predictor under PS, even though these parameters may not coincide with the true latent random field. 

\section{Inverse sampling intensity weighting (ISIW)}\label{sec:aim2_ISIW}
We now describe the ISIW procedure for PS adjustment. ISIW is a two-stage approach, wherein we first estimate the sampling intensity at each of the observation locations $\textbf{X}$. In the second stage, we input the (estimated) inverse sampling intensities as weights into a weighted likelihood adjustment for a chosen composite likelihood. For spatial interpolation, the resulting adjusted parameter estimates are then substituted in the kriging equations in \eqref{eqn:aim2_krigall}.  
 
\subsection{Sampling intensity estimation}
The essence of ISIW is to account for the dependence between the response and observation locations by using the vector of estimated sampling intensities instead of the full likelihood of the observation process. The heuristic motivation for ISIW is to reweight observations by inverse sampling intensity to approximate a representative, non-preferential spatial sample of the study region.

Methods to estimate a spatially varying intensity of a point process can be divided into parametric and nonparametric approaches. Domain knowledge can inform the parametric form of the intensity either through choice of model or covariates, but oftentimes this information is unavailable and nonparametric estimation is required. The nonparametric approaches follow the kernel smoothing approach discussed in \cite{diggle_kernel_1985} which we will refer to as kernel intensity estimators (KIE). Let $k$ be a $d$-dimensional kernel function from $\mathbb{R}^d \to \mathbb{R}^+$, which is a symmetric probability density function. Given a bandwidth size $h>0$ and edge correction factor $w_h(\cdot, \cdot)$, the KIE estimate at a point $\mathbf{x}$ is given by
\begin{equation*}
    \hat{\lambda}(\textbf{x};h)=h^{-d} \sum_{\textbf{s} \in \textbf{X}\cap \mathcal{D}} k\left(\frac{\lVert\textbf{x} - \textbf{s}\rVert}{h}\right) w_h(\textbf{x},\textbf{s})^{-1}, \textbf{x}\in \mathcal{D}.   
\end{equation*}

The key hyperparameter for KIE is the bandwidth size $h$. Bandwidth selection is a well-studied problem, with methods ranging from high bandwidth, smooth intensity estimators to low bandwidth, flexible intensity estimators. Each has its place in the bias-variance tradeoff, with higher bandwidths exhibiting higher bias with lower variance and lower bandwidths vice versa. We experiment with several bandwidth selection strategies implemented in the \texttt{spatstat} \texttt{R} package (Table \ref{tab:aim2_bw_selection}).

While KIEs provide valid estimates of the first-order intensity based solely on a realization of the point pattern, a potential limitation for PS adjustment is that they do not incorporate information from the observed responses $\mathbf{Y}$. Under the SLP model in \eqref{eqn:slp}, for example, the sampling intensity depends on the latent process, which also drives the observed response values. Incorporating both response values and spatial locations may therefore yield improved estimates of the sampling intensity relative to approaches based only on the locations.

To explore this idea, we study an extension of KIE that incorporates the response through a spatial covariate using the kernel-based approach of \citet{baddeley_nonparametric_2012}. We also considered a random forest–based intensity estimation method proposed by \citet{biscio_nonparametric_2025}; however, its results were very similar to those obtained with KIE so we do not present them here. Specifically, we use a predicted version of $\mathbf{Y}$ as an additional covariate in the intensity estimation. Because KIEs with spatial covariates require covariate values to be available at all grid locations, we obtain this covariate by kriging using parameters estimated via MLE. Although this predicted surface is biased, as has been extensively discussed, PS induces systematic discrepancies between the true process and the estimated surface. If such structure is present, the nonparametric relationship between the covariate and the sampling intensity may still be learnable. We refer to this approach as ``ISIW KIE COV" and the approach using just the locations as ``ISIW KIE." 

As with other inverse weighting procedures, a key concern for ISIW is the presence of extreme weights. Observation locations in sparsely sampled regions, which occur more frequently under PS designs, can receive very large weights in the resulting likelihood and may lead to numerical instability during optimization. Two commonly used approaches from the inverse probability weighting literature are \textit{trimming} and \textit{winsorization}. Trimming removes observations deemed to be outliers, whereas winsorization preserves the sample size by capping extreme values at a specified threshold. 

We adopt winsorization to mitigate the impact of extreme weights, as trimming observations may discard too many informative locations, thereby weakening the effectiveness of the PS adjustment. Our winsorization procedure proceeds as follows. First, we estimate the sampling intensities and take their inverse. An upper quantile is then selected as the winsorization threshold, and all values exceeding this threshold are set equal to it. Finally, the weights are normalized to sum to the total number of observations \(n\). Denote the $\tau$-level empirical quantile of the estimated inverse sampling intensities $\hat{\lambda}^{-1}$ as $W_\tau$. The resulting estimated weight for the \(i\)th observation is
\begin{equation} \label{eqn:aim2_weights}
    \hat{w}_i
    =
    n \cdot
    \frac{\min\{\hat{\lambda}(\mathbf{x}_i)^{-1}, W_\tau \}}
    {\sum_{\mathbf{s} \in \mathbf{X}} 
\min\{ \hat{\lambda}(\mathbf{s})^{-1}, W_\tau \}} .
\end{equation}

\subsection{Weighted likelihood and estimation}
ISIW applies weights proportional to the inverse sampling intensity to each event in a likelihood factored as a product of densities. While such weighting is straightforward in settings with independent observations, the appropriate weighting scheme is less clear in the presence of spatial dependence and correlation among observations. Existing applications of ISIW in the PS literature include univariate marginal \citep{vedensky_look_2023} and pairwise difference CLs \citep{schliep_correcting_2023}. These approaches yield intuitive weighting schemes by enforcing independence across subsets of events in the joint density. For reference, the weighted univariate marginal CL follows
\begin{equation*} \label{isiw-um}
\begin{split}
    \log \mathcal{L}_{WM}(\mu;\textbf{y}) = \sum_{i=1}^n w_i\log f(y_i;\mu),
\end{split}
\end{equation*}
while the pairwise difference CL follows
\begin{equation*} \label{isiw-pm}
\begin{split}
    \log \mathcal{L}_{WD}(\theta;\textbf{y}) = \sum_{i=1}^n\sum_{j=i+1}^n w_i w_j \log f(y_i- y_j;\theta).
\end{split}
\end{equation*}
However, the univariate marginal CL cannot estimate covariance parameters $\theta$ while the pairwise difference CL cannot estimate $\mu$ directly, limiting its impact on PS adjustment. In addition, both approaches ignore higher-order spatial dependence, which is essential for accurately approximating Gaussian process models. 

To address these gaps, we explore the application of ISIW to the Vecchia approximation. Let the weights be defined as in \eqref{eqn:aim2_weights}. We initially considered weighting the likelihood as
\begin{equation*} \label{isiw-v-1}
\begin{split}
    \log \mathcal{L}_{WV}(\boldsymbol{\psi};\textbf{y})&=w_{p(1)}\log f(y_{p(1)};\boldsymbol{\psi})\\
    &+\sum_{i=2}^n \left(\prod_{j\in \{p(i)\}\cup q(i)} w_j \right)\log f(y_{p(i)},\textbf{y}_{q(i)};\boldsymbol{\psi}) \\ 
    &- \sum_{i=2}^n \left(\prod_{j\in q(i)} w_j  \right)\log f(\textbf{y}_{q(i)};\boldsymbol{\psi}),
\end{split}
\end{equation*}
to maintain the probabilistic interpretation of the inverse weighting for the component conditional densities. However, numerical issues arise due to taking the product of several weights, greatly increasing the chance of extreme values. As a more stable approach, we approximate the true weight using 
\begin{equation} \label{isiw-v-2}
\begin{split}
    \log \mathcal{L}_{WV}(\boldsymbol{\psi};\textbf{y})&=w_{p(1)}\log f(y_{p(1)};\boldsymbol{\psi})\\
      &+\sum_{i=2}^n w_{p(i)}\log f(y_{p(i)},\textbf{y}_{q(i)};\boldsymbol{\psi}) \\ 
    &- \sum_{i=2}^n w_{p(i)}\log f(\textbf{y}_{q(i)};\boldsymbol{\psi}).
\end{split}
\end{equation}
We also considered the pairwise marginal composite likelihood as a candidate for ISIW, as it naturally accommodates weighting while allowing estimation of $\mu$. However, preliminary analyses indicated that ISIW applied to the Vecchia approximation outperformed the pairwise marginal CL across all evaluation metrics (Table \ref{tab:aim2_rank}). We therefore focus the remainder of the paper on ISIW combined with Vecchia.

The ISIW parameters can be numerically estimated by standard optimization procedures. We used the L-BFGS-B routine as implemented in the \texttt{optim} package in the \texttt{R} language. In our experience, other derivative-free optimization procedures such as the Nelder-Mead algorithm also work well. Initial values for parameters to be estimated were selected with general rules of thumb from the \texttt{GPVecchia} R package and all other optimization parameters were set to their default settings. 

\subsection{Prediction}
Although ISIW involves parameter estimation, the estimated parameters primarily serve as a means for generating probabilistic predictions through the kriging equations. It is well known that under standard infill asymptotics, the Gaussian process parameters $\mu$, $\phi$, and $\sigma^2$ in a Mat\'ern covariance model are weakly identified and cannot be estimated consistently \citep{zhang_inconsistent_2004, wang_prediction_2020}. Accordingly, ISIW methods are better suited for improving probabilistic prediction under PS than for recovering the true model parameters. Nevertheless an effective CL should aim to estimate as many GP parameters as possible in order to maximize the potential for PS adjustment through weighting. In this regard, the Vecchia approximation estimates the full set of mean and covariance parameters, in contrast to univariate marginal and pairwise difference CLs, which is expected to yield improved predictive performance.

Prediction under ISIW follows by substituting $\hat{\boldsymbol\psi}$ obtained by maximizing \eqref{isiw-v-2} into the kriging equations in \eqref{eqn:aim2_krigall}. This is a key distinction between ISIW and the SLP. Whereas prediction by ISIW adheres to kriging by substituting PS-adjusted parameters, the SLP model computes predictions from the estimated distribution of $[S\mid \textbf{X, Y}]$, making it much more computationally intensive. 

Prediction variance for ISIW is estimated using the ordinary kriging variance formula \citep{cressie_statistics_2015}, which adds a penalty term to the simple kriging variance in \eqref{eqn:aim2_krigall} for estimating $\mu$. The variance is computed by substituting estimates of $\theta$ from maximizing \eqref{isiw-v-2} into
\begin{equation*}
\begin{split}
\sigma^2_{WV}(\mathbf{X}_0) &= C_\theta(\textbf{X}_0, \textbf{X}_0) - C_\theta(\textbf{X}_0, \textbf{X}_n)^\top\boldsymbol\Sigma_n^{-1}(\theta) C_\theta(\textbf{X}_0, \textbf{X}_n) \\&+  
\frac{\bigl(
1
-
\mathbf{1}^\top
{\Sigma}_n({\theta})^{-1}
{C}_{{\theta}}(\mathbf{X}_0, \mathbf{X}_n)
\bigr)^2}
{\mathbf{1}^\top
{\Sigma}_n({\theta})^{-1}
\mathbf{1}} .
\end{split}
\end{equation*}
\cite{schliep_correcting_2023} propose kriging variance estimators under PS, however their method does not apply to our scenario because they seek to recover the variance had the population been sampled. In contrast, our goal is to estimate prediction variance as a quantification of uncertainty, rather than to recover an estimate from a superpopulation.

\section{Simulation analysis}\label{sec:aim2_simulation}

In this section, we present a series of simulation experiments designed to evaluate the performance of ISIW. We first describe the evaluation metrics and preliminary analyses used to select a representative KIE from a set of candidates as well as the weight threshold for the winsorization. We then conduct describe various simulations comparing ISIW KIE, ISIW KIE COV, and ISIW with the true sampling intensities (ISIW Known, which serves as an oracle implementation) against the MLE and SLP across a range of random field specifications. We assess robustness to model misspecification and characterize conditions under which each method performs well for parameter estimation and prediction.

\subsection{Evaluation metrics}

To evaluate the probabilistic predictions of our method, we used the \textit{continuous ranked probability score} (CRPS) to compare predictions over a test grid while accounting for prediction variance \citep{gneiting_strictly_2007}. When the predictive distribution is Gaussian, a closed form formula for the CRPS score is
\begin{equation*}
\mathrm{CRPS}
= \frac{1}{n} \sum_{i=1}^n 
\sigma_i \left[
z_i \left( 2\Phi(z_i) - 1 \right)
+ 2\phi(z_i)
- \frac{1}{\sqrt{\pi}}
\right], \qquad 
z_i = \frac{y_i - \hat{y}_i}{\sigma_i},
\end{equation*}
where $\Phi(\cdot)$ and $\phi(\cdot)$ are the standard normal cumulative distribution function and probability density function, respectively; $\sigma_i$ is the square root of the prediction variance at test location $\mathbf{x}_i$, $y_i$ is the true value, and $\hat{y}_i$ is the corresponding point prediction. As an alternative measure, we also computed \textit{root mean squared prediction error} (RMSPE). To compare parameter estimates, we evaluated bias and root mean squared error (RMSE) relative to the true values.

\subsection{Preliminary analyses}

We first picked a representative composite likelihood and KIE to use in the main simulation study. Specifically, we considered the pairwise marginal CL and Vecchia approximation, combined with several KIE choices listed in Table \ref{tab:aim2_bw_selection}. Based on a preliminary analysis ranking all combinations by RMSPE over a variety of random field scenarios (Supplementary Section \ref{sec:KIE_choice}) we chose the Vecchia approximation combined with the \texttt{CvL.adaptive} KIE bandwidth selection criterion \citep{cronie_non-model-based_2018, van_lieshout_infill_2021} which consistently achieved the highest RMSPE ranking (Table \ref{tab:aim2_rank}).  

After selecting an ISIW implementation, we determined the winsorization threshold through a separate preliminary analysis conducted under similar settings to those used in the main simulation experiment (Supplementary Section \ref{sec:Winsorization_choice}). Based on a sensitivity analysis evaluating predictive performance across candidate quantiles (Tables \ref{tab:winsorization_sensitivity_rmspe}, \ref{tab:winsorization_sensitivity_crps}), we winsorized extreme weights to the 93\% percentile for all later analyses. Winsorization was only applied to weights for ISIW KIE and ISIW KIE COV and not ISIW Known.

\subsection{Point process misspecification}

We generated $B=500$ realizations for two separate Mat\'ern random fields on a $200\times200$ grid on the unit square: a low-range ($\phi=0.02$) and high-range ($\phi=0.15$) surface with common parameters $\mu=4, \sigma^2=1.5, \nu=1,$ and $\tau^2=0.1$. For each realization, we sampled observation locations according to two distinct point process models with $n=100$: a log Gaussian Cox process (LGCP) and a Thomas process (a parent-offspring point process). Traditionally, the SLP has only been evaluated for LGCP designs, where it is correctly specified. A small sample size was chosen to align closer with the finite samples encountered in our real data applications.

For the LGCP, the sampling intensity was given by $\exp\{\beta S(\mathbf{x})\}$. For the Thomas process, parent points were first generated from a homogeneous Poisson process with an expected count of $300$. The number of offspring for each parent then followed a Poisson distribution with mean $\exp\{\beta S(\mathbf{x})\}$ and offspring points were sampled from a normal distribution centered at their respective parent locations, with a scale parameter of $0.03$. The resulting point pattern was conditioned to contain $n=100$ points, just like the LGCP. We considered values of $\beta\in\{-1, 1,2\}$ to investigate differences under varying strengths of PS. 

% Describe benchmarks and estimation procedures
For each scenario, we fit the MLE, SLP, ISIW KIE, ISIW KIE COV, and ISIW Known methods. We estimated the SLP model parameters using the INLA-SPDE implementation as defined in \cite{watson_general_2019} and used penalized complexity (PC) priors on the variance and range parameters \citep{simpson_penalising_2017, fuglstad_constructing_2019}. We adopted an empirical Bayes approach, specifying priors such that $P(\phi<\phi_0)=0.50$ and $P(\sigma>\sigma_{0})=0.01$ where $\phi_0$ is one quarter of the mean pairwise distance among sampled locations and $\sigma_0=\sqrt{0.9\cdot \widehat{\text{Var}}(Y)}$ where $\widehat{\text{Var}}(Y)$ denotes the sample variance of the observed responses. The default priors for INLA were used for $\mu$ and $\tau$, set as $N(0, 1000)$ and $\text{Gamma}(1, 5\times 10^{-5})$, respectively. We will refer to the SLP as the INLA-SLP. Prediction was evaluated by computing the mean CRPS and RMSPE over the centroids of a $32 \times 32$ discretization of the unit square. Parameter estimation was assessed by computing bias and RMSE for estimating $\boldsymbol{\psi}:=(\mu, \sigma^2, \phi, \tau^2)$. For all methods, $\nu$ was assumed to be fixed. 

Prediction results appear in Table \ref{tab:synthetic_crps}. Across all methods, prediction error increased with stronger PS (larger $|\beta|$) and weaker spatial correlation (smaller $\phi$). When the sampling design was correctly specified as a LGCP, INLA-SLP consistently achieved the lowest CRPS across all values of $\beta$ and $\phi$. In contrast, under the misspecified Thomas process, ISIW Known yielded the best predictive performance across all scenarios. In general, ISIW Known and ISIW KIE predicted better than the MLE, demonstrating the benefit of ISIW for spatial interpolation under PS. However, the CRPS for ISIW Known was noticeably lower than that of both ISIW KIE and ISIW KIE COV, indicating a gap between the information contained in the known versus estimated weights. In particular, ISIW KIE COV generally underperformed, with CRPS values comparable to the MLE and, in several low range ($\phi=0.02$) scenarios, worse than the MLE. The largest gains of ISIW Known compared to INLA-SLP occurred under $\beta=2$ in the misspecified Thomas setting, while ISIW KIE also outperformed INLA-SLP for the Thomas process at $\beta=2$ and performed competitively when $\phi=0.15$ for $\beta=-1$ and $\beta=1$. 

\begin{figure}[!ht]
\centering
\includegraphics[]{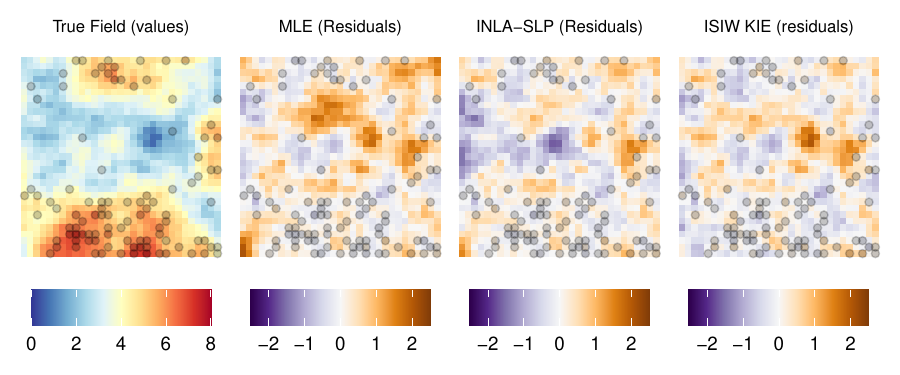}
\caption{Example predictive surface for the MLE, INLA-SLP, and ISIW KIE approaches for a LGCP point pattern. The first panel shows the true field values with observations denoted as points, while the remaining panels display residuals (predicted minus truth) for each method.}
\label{fig:aim2_prediction_example}
\end{figure}

Figure \ref{fig:aim2_prediction_example} provides an illustrative example comparing the predictive tendencies of MLE, INLA-SLP, and ISIW Known. In the data-rich area at the bottom, predictions from all three methods are nearly identical. The differences emerge in data-sparse regions where PS is known to have its greatest impact. While the MLE overpredicts the center of the grid, the INLA-SLP predicts a much lower value over the same region which decreases predictive error but leads to underestimation in a specific area indicated by the deep purple. ISIW, similar to INLA-SLP, lowers predictions in that region but not as drastically, avoiding the central underprediction seen in INLA-SLP but still overestimating in another area, shown as dark orange slightly to the right of the center. As shown by this example, both INLA-SLP and ISIW adjust for PS in similar ways, but INLA-SLP tends to apply a larger adjustment in data sparse areas compared to ISIW.

Table \ref{tab:synthetic_mu_bias_rmse} summarizes bias and RMSE for estimation of $\mu$. The MLE exhibited substantial bias and RMSE across all settings, with error increasing with larger $|\beta|$ and smaller $\phi$. In low range settings, ISIW Known generally achieved the smallest error, whereas INLA-SLP performed best in high range settings across both LGCP and Thomas designs. The largest gains of ISIW Known relative to INLA-SLP occurred in low range scenarios. ISIW KIE had higher estimation error than INLA-SLP but lower than the MLE, while ISIW KIE COV showed large bias and RMSE overall, consistent with its poor prediction. Notably, INLA-SLP had the worst performance under the Thomas process with $\phi=0.02$ and $\beta=2$, coinciding with severe underestimation of $\mu$, underscoring the importance of accurate mean estimation for prediction under PS in finite samples.

We now summarize results for covariance parameter estimation (Tables \ref{tab:sigma2_bias_rmse}, \ref{tab:range_bias_rmse}, \ref{tab:nugget_bias_rmse}). Overall, trends in covariance parameter estimation were less distinct than those observed for mean estimation, and their relationship with prediction is less clear. The MLE was generally the most stable and reliable method for estimating the range and nugget parameters. Error for the range was lower when $\phi=0.02$ while error for the nugget was lower when $\phi=0.15$. In fact, the nugget was well estimated by all methods when $\phi=0.15$, although the MLE consistently achieved the lowest bias and RMSE. For variance estimation, INLA-SLP tended to perform well when $\phi=0.15$ while the MLE performed best when $\phi=0.02$, suggesting a broader trend that INLA-SLP performs better in prediction and inference under stronger spatial correlation.

However, we found that INLA-SLP was numerical unstable in some low range simulations leading to inflated estimates of covariance parameters, with isolated instances even when $\phi=0.15$. None of these numerical issues occurred in the commonly studied scenario of $\beta=1$, $\phi=0.15$ and LGCP design, indicating that the simulation setting traditionally used to evaluate INLA-SLP works quite well. ISIW methods generally showed larger error for covariance parameters than both MLE and INLA-SLP and was also numerically unstable in low range scenarios. Despite these large estimates for the variance or range, prediction did not suffer too much as evidenced by Table \ref{tab:synthetic_crps}. These results indicate that low prediction error under PS does not necessarily correspond to accurate covariance parameter estimation compared to accurate estimation of $\mu$. 

\subsection{Random field misspecification}

In the previous simulation experiment, we demonstrated how misspecification of the sampling design \(\mathbf{X}\) can degrade the performance of INLA-SLP and highlighted the benefit of ISIW, which estimates the sampling intensity nonparametrically. We now consider a setting in which the latent random field itself is misspecified, being composed of a sum of multiple random fields representing different components of the underlying process rather than a single Gaussian process with Mat\'ern covariance. Because parameter estimation no longer corresponds to the true data-generating mechanism in this setting, we focus exclusively on predictive performance as measured by the CRPS.

We generated \(B=500\) realizations of a latent field \(S\) defined as
\begin{equation*}
    S(\mathbf{x}) = S_\ell(\mathbf{x}) + S_h(\mathbf{x}) + S_m(\mathbf{x}),
\end{equation*}
where \(S_\ell\), \(S_h\), and \(S_m\) are independent Gaussian processes. In this construction, \(S_\ell\) corresponds to a \textit{low}-range GP, \(S_h\) to a \textit{high}-range GP, and \(S_m\) to a \textit{medium}-range GP that serves as the shared latent variable governing the sampling intensity. Each process has mean zero and marginal variance \(\sigma^2 = 0.5\). For \(S_\ell\), we set \(\phi = 0.02\) and \(\nu = 1/2\); for \(S_h\), \(\phi = 0.30\) and \(\nu = 2\); and for \(S_m\), \(\phi = 0.15\) and \(\nu = 1\). This construction induces heterogeneous smoothness and spatial dependence across scales and generates a latent field that cannot be represented by a single Mat\'ern GP. A nugget effect with variance \(\tau^2 = 0.05\) was added to each realization of \(S\).

Observation locations were then sampled according to an intensity function of the form \(\exp\{\beta S_m(\mathbf{x})\}\) with \(\beta = 1.5\). Both LGCP and Thomas point process designs were considered, defined as in the earlier synthetic data experiment, with a fixed sample size of \(n=100\). Results of the experiment are summarized in Table~\ref{tab:synthetic_multipleGP_crps}. When the sampling design was LGCP, INLA-SLP and ISIW Known achieved nearly identical CRPS values of 0.508 and 0.507, respectively. This contrasts with the earlier simulation experiment, in which INLA-SLP consistently outperformed ISIW Known by a substantial margin under the LGCP design. Under the misspecified Thomas design, INLA-SLP predicted the worst (CRPS = 0.637), while ISIW methods provide substantial improvements. ISIW KIE predicted only marginally better than the MLE, and ISIW KIE COV again underperformed relative to the other ISIW variants.

\subsection{Real data simulation}

To better assess the behavior of ISIW in settings resembling our real data applications, we generated random fields and sampled observations calibrated to match the properties of the Galicia and California AQS datasets, with parameters obtained from INLA-SLP fits to the observed data. We first scaled all Easting and Northing coordinates by 100,000 meters. For the Galicia data, we sampled $n=63$ points from a SLP with $\mu=2.18$, $\sigma^2=0.146$, $\nu=1$, $\phi=0.838$, $\tau^2=0.193$, and $\beta=-5.20$ using the same bounding box as the original study region. For the California AQS data, we sampled $n=98$ points from a SLP with $\mu=0.88$, $\sigma^2=0.567$, $\nu=1$, $\phi=1.15$, $\tau^2=0.016$, and $\beta=2.47$.

Although alternative sampling designs besides the LGCP could be considered, we have already demonstrated that misspecification of either component can affect predictive performance of INLA-SLP. The goal of this experiment is therefore to evaluate whether ISIW can still outpredict MLE even when the data are generated from a SLP model derived from the real data. Results of the experiment are reported in Table~\ref{tab:semi_synthetic_crps}. Across both regions, all ISIW variants improved predictions relative to the baseline MLE, with ISIW Known achieving the lowest CRPS, followed by ISIW KIE and ISIW KIE COV. We noticed larger improvements in prediction error for the ISIW in the Galicia setup compared to the California setup.

\section{Real data analysis} \label{sec:aim2_galicia}

\subsection{Galicia moss biomonitoring}

The Galicia moss dataset has been a widely used example for illustrating the effects of PS on geostatistical inference and prediction \citep{diggle_geostatistical_2010, dinsdale_methods_2019, silva_exact_2024}. It contains 63 measurements from 1997 and 132 measurements from 2000 of lead concentrations in moss samples, measured in micrograms per gram of dry weight, collected from Galicia, northern Spain (Figure \ref{fig:aim2_galicia_data}). Sampling in 1997 was preferential, with a bias toward locations in the north with lower lead concentrations, whereas sampling in 2000 was more regular and non-preferential. We fit the MLE, INLA-SLP, ISIW KIE and ISIW KIE COV on the 1997 log-transformed data and generated predictive surfaces over a $35 \times 35$ grid covering the Galicia area. While previous analyses of the Galicia dataset have used areal models, we elected to use the SPDE approach for INLA-SLP to be consistent with our simulation analysis.

\begin{figure}[h]
\centering
\includegraphics[width=\textwidth]{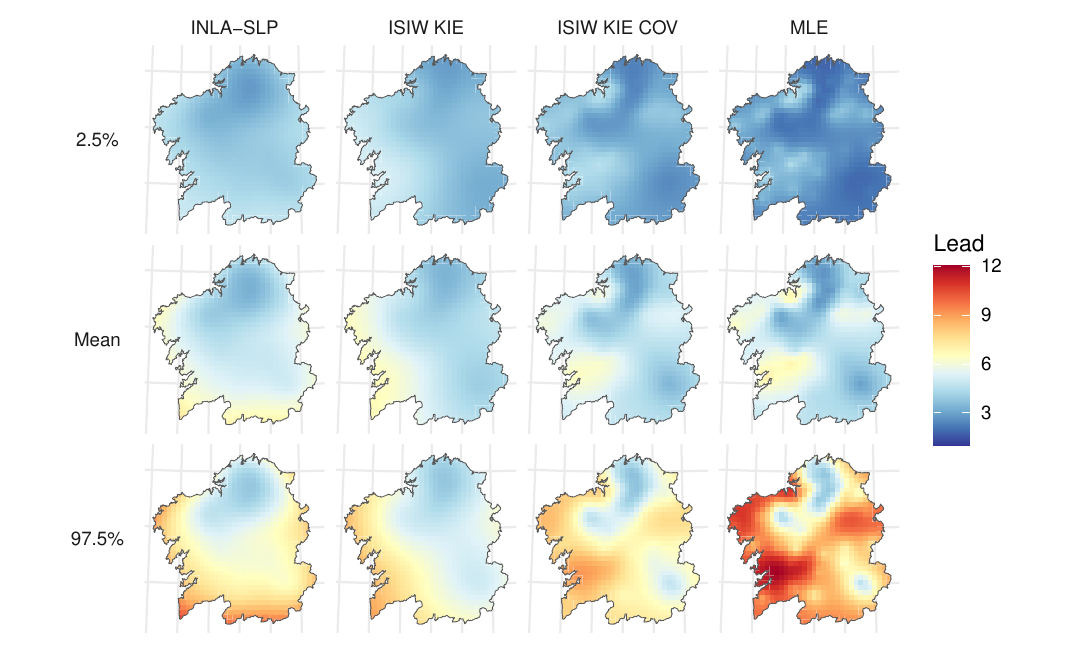}
\caption{Spatial predictions and uncertainty of lead concentrations in Galicia.}
\label{fig:galicia_pred}
\end{figure}

\begin{figure}[!ht]
\centering
\includegraphics[width=\textwidth]{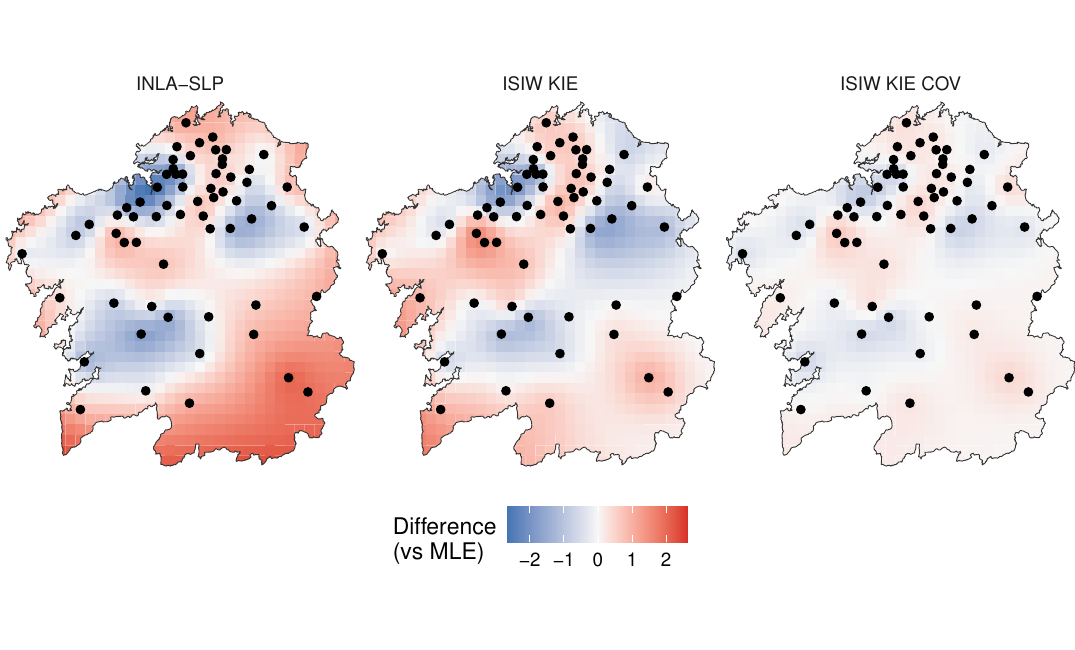}
\caption{Differences in point predictions between each method and the MLE for the Galicia data. Black points represent observed locations.}
\label{fig:galicia_diff}
\end{figure}

Figure \ref{fig:galicia_pred} displays predictive surfaces and corresponding lower and upper quantiles under MLE, INLA-SLP, ISIW KIE, and ISIW KIE COV. Across methods, the broad spatial patterns were similar, with lower predicted concentrations in the north and higher levels toward the south and west. At a finer scale, MLE and ISIW KIE COV predicted greater spatial heterogeneity driven by the observed data, whereas INLA-SLP and ISIW KIE produced smoother predictive surfaces. All three PS methods (INLA-SLP, ISIW KIE, and ISIW KIE COV) were spatially consistent in the direction of adjustment relative to the MLE (Figure \ref{fig:galicia_diff}), differing primarily in the magnitude of correction. INLA-SLP produced the largest deviations from the MLE, followed by ISIW KIE and ISIW KIE COV, consistent with the behavior illustrated in the simulation from Figure \ref{fig:aim2_prediction_example}.

% Figure \ref{fig:aim2_galicia_pred_isiw} displays the predictions for the log lead concentration surface for MLE, ISIW-V and ISIW-PM. Figure \ref{fig:aim2_galicia_pred_inla} displays the same for the INLA-SLP under the three variance parameter priors. Substantial differences emerged in the 1997 data. As expected, the MLE predicts low in data sparse areas due to PS while both the ISIW methods and INLA-SLP estimated higher values in data sparse areas to correct for the preferential sampling of low values. When the PC prior for INLA-SLP was set to $P(\sigma^2>1)=0.01$, the predictive surface appears reasonable. However, for $P(\sigma^2>0.5)=0.01$ the INLA-SLP corrects for PS in data sparse areas by an unreasonably high amount, whereas for $P(\sigma^2>0.1)=0.01$, the predictive surface becomes so smooth that it does not pick up on any of the variation in the observations. INLA-SLP can be highly sensitive to small changes in prior specification, and careful analysis is necessary for correct implementation. This is even more evident by how much the parameter estimates of INLA-SLP change under each prior (Table \ref{tab:galicia_param}). ISIW-V appears to strike a balance between implementing a moderate adjustment for PS while preserving the spatial correlation evident in the dataset. 

\subsection{2010 California Air Quality System (AQS) monitoring}

\begin{figure}[!ht]
\centering
\includegraphics[scale=.95]{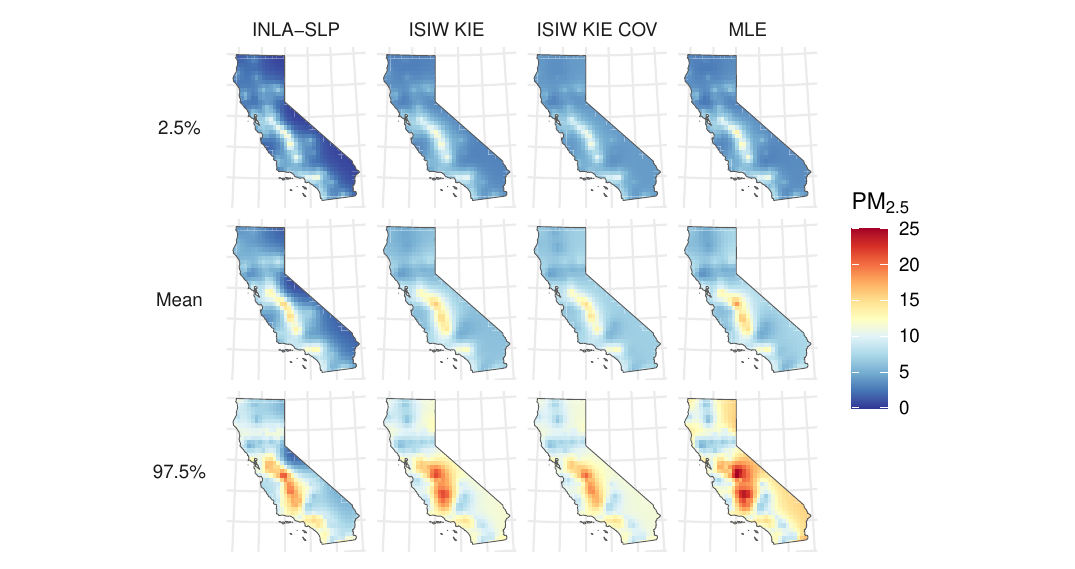}
\caption{Spatial predictions and uncertainty of air pollution concentrations in California.}
\label{fig:california_pred}
\end{figure}

\begin{figure}[!ht]
\centering
\includegraphics[width=\textwidth]{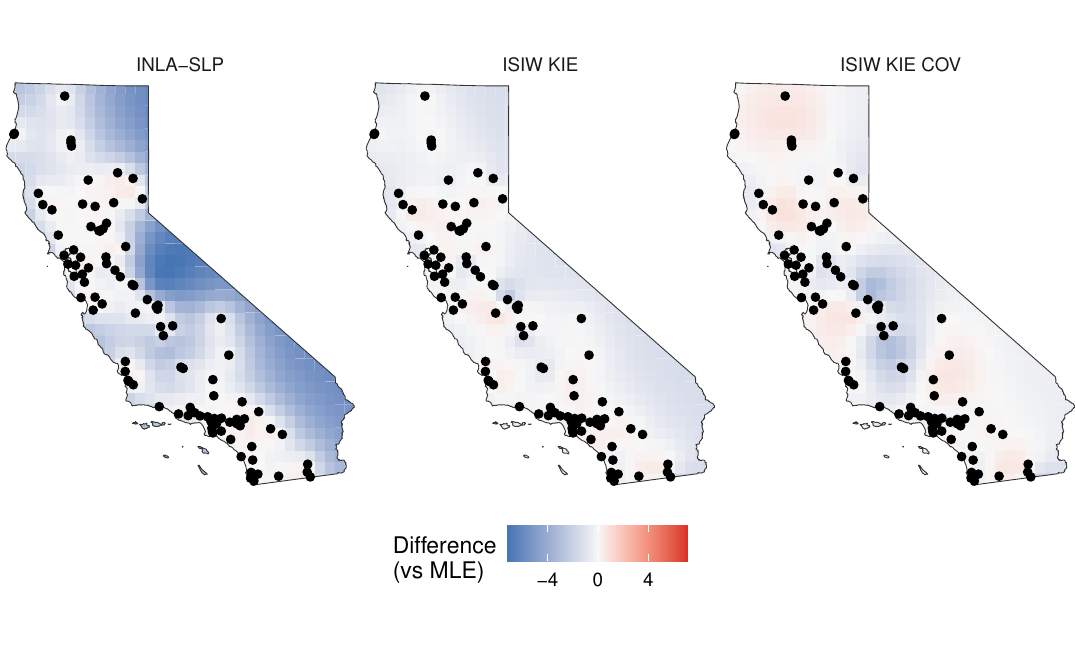}
\caption{Differences in point predictions between each method and the MLE for the California AQS data. Black points represent observed locations.}
\label{fig:california_diff}
\end{figure}

We estimated average daily $\text{PM}_{2.5}$ concentrations in 2010 using data from 98 EPA AQS monitoring locations in California (Figure \ref{fig:ca_aqs_data}). Figure~\ref{fig:california_pred} presents the predictions, with differences relative to the MLE shown in Figure~\ref{fig:california_diff}. All methods once again had similar spatial patterns, with lower concentrations in northern and coastal regions and higher concentrations in the Central Valley and parts of southern California. Estimated concentrations generally ranged from approximately 5--10~$\mu\text{g}/\text{m}^3$ in northern California to 12--18~$\mu\text{g}/\text{m}^3$ in the Central Valley, with higher values in the upper tail of the predictive distribution.

INLA-SLP produced the largest departures from the MLE, yielding lower mean predictions across much of northern California, the eastern Sierra Nevada, and interior southern California, with differences typically ranging from approximately $5$--$7~\mu\text{g}/\text{m}^3$ in sparse data areas. ISIW KIE also tended to estimate lower concentrations than the MLE over much of the state, particularly in inland areas, though these differences were minimal, generally within $1$--$2~\mu\text{g}/\text{m}^3$. ISIW KIE COV additionally predicted higher than MLE in parts of the Central Valley and coastal California, with increases on the order of 1--3~$\mu\text{g}/\text{m}^3$. In general, ISIW produced very similar predictions to the MLE relative to the Galicia application, while INLA-SLP made a much more noticeable adjustment for PS. This echoes the results from the real data simulation experiment.

\section{Discussion} \label{sec:aim2_discussion}
% Emphasize novelty
In this work, we investigated ISIW as a practical approach for adjusting PS in model-based geostatistics. By combining ISIW weights estimated by KIE with the Vecchia approximation, we developed a computationally efficient method that improves spatial prediction under PS while avoiding the need to fully specify the likelihood of the observation process $\mathbf{X}$. Across a broad set of simulation experiments, ISIW outpredicted the MLE and even the SLP model when the sampling design was misspecified. 

The main finding of this study is that ISIW improves spatial prediction across a wide range of PS scenarios. These scenarios include both positive and negative PS effects, alternative sampling designs, increasing strengths of PS, and different latent random field structures, including additive Gaussian processes and processes calibrated to resemble our real data applications. In particular, ISIW outpredicted SLP when the design followed a non-LGCP pattern, highlighting the sensitivity of the SLP to point process misspecification. In contrast, ISIW relies on a nonparametrically estimated sampling intensity, and was shown to still predict well when the true sampling mechanism deviated from standard SLP assumptions. Robustness is key for environmental and ecological applications, where sampling designs are often driven by operational constraints and may contain complex clustering that is difficult to emulate using standard random fields. 

Combining ISIW with the Vecchia approximation has clear advantages over earlier pairwise and univariate CL approaches. Whereas CL reduces computation by discarding much of the joint dependence in the data, the Vecchia approximation captures higher order correlations while remaining scalable. At the same time, the resulting weighting scheme is less intuitive in the Vecchia likelihood and is sensitive to ordering and tuning decisions such as winsorization thresholds, underscoring the need for more guidance on hyperparameter selection in future work.

Another key finding from our simulation experiment was the disconnect between parameter estimation accuracy and predictive performance under PS. In settings with weak spatial dependence, ISIW achieved low error in estimating $\mu$, consistent with theoretical results for weighted likelihood under independence. However, estimation was more challenging in the high range setting, implying this theoretical justification weakens under spatial correlation. We did not observe analogous results in ISIW covariance parameter estimation, which was highly biased overall, whereas the MLE consistently achieved the lowest covariance parameter estimation error. Even so, ISIW frequently predicted better than the MLE. This suggests that in finite samples accurate estimation of covariance parameters is not essential for effective spatial prediction under PS compared to that of $\mu$. In addition, kriging-based predictors, while not optimal, can perform well under PS even when the fitted model does not correspond to the true generative model.

Our work also revealed some limitations of the traditional SLP model. While INLA-SLP performed well when the LGCP assumption was correct and spatial dependence was strong, it exhibited numerical instability in covariance parameter estimation under low range conditions and was sensitive to misspecification of the sampling design. INLA-SLP also tended to apply more extreme adjustments in data-sparse regions than ISIW, sometimes leading to overcorrection. Prior specification and sensitivity analyses remain large standing issues for running complex INLA models (Figure \ref{fig:aim2_galicia_pred_inla}). These properties underscore the importance of evaluating PS methods under a broader range of data-generating mechanisms than those typically considered in the literature.

ISIW may also be useful in other settings where a latent GP drives the preferential sampling of locations, such as species distribution modeling (SDM) and ecological studies based on presence–absence or count data. In these applications, sampling effort is often spatially structured and correlated with the underlying intensity of the process of interest, leading to biased inference and prediction if ignored. While existing approaches in these domains typically rely on binary or count-based models, our work focuses on PS adjustment for continuously observed Gaussian outcomes. Nonetheless, the core idea of weighting likelihood contributions to account for PS is broadly applicable, suggesting that extensions of ISIW to SDMs may offer a promising direction for future research.

The gap between ISIW Known and ISIW with estimated weights indicates that nonparametric intensity estimation remains an area of improvement. Our simulations suggest that in finite samples KIEs can recover some but not all of the information needed for optimal adjustment. The ISIW KIE COV variant, which incorporates a biased predicted response surface as a spatial covariate in the intensity estimation, did not perform as well as expected. While this approach was intentionally exploratory, the results suggest that naïvely incorporating response information into intensity estimation is insufficient. More principled methods that integrate response values without requiring full joint likelihood specification may push ISIW KIE performance closer to that of ISIW Known.

Uncertainty quantification under ISIW remains a challenge. We relied on the ordinary kriging variance formula using PS adjusted parameter estimates as a working measure of predictive uncertainty. While this approach is pragmatic and likely conservative, and simulations showed that ISIW predicts well compared to the other methods, its coverage properties under PS remain unclear. Bootstrap approaches could provide improved uncertainty quantification, but are computationally demanding and require resampling spatial point patterns for which principled procedures are not as well-established. Developing methods for ISIW prediction variance is an important direction for future work.

Despite these limitations, ISIW offers several practical advantages that make it appealing for applied use. It is computationally fast, conceptually simple, and easy to integrate with modern Gaussian process approximation methods, many of which already rely on Vecchia-type factorizations. Unlike the SLP, ISIW avoids the need to specify and estimate a joint likelihood for the response and sampling locations, reducing both modeling complexity and computational burden. 

Overall, this work suggests that improved spatial prediction under PS can be achieved without full joint modeling of the observation and response processes. Further advances in intensity estimation for marked point processes and uncertainty quantification have the potential to enhance ISIW even further for geostatistical analysis under PS.

\section*{Ethics declaration}
\subsection*{Conflict of interest}
The authors have no conflict of interest to declare.

\newpage
\pagestyle{empty}

\bibliographystyle{apalike}  % Author-year citation format
\bibliography{references}  % Ensure you have a references.bib file

@article{gneiting_strictly_2007,
	title = {Strictly {Proper} {Scoring} {Rules}, {Prediction}, and {Estimation}},
	volume = {102},
	issn = {0162-1459, 1537-274X},
	url = {http://www.tandfonline.com/doi/abs/10.1198/016214506000001437},
	doi = {10.1198/016214506000001437},
	language = {en},
	number = {477},
	urldate = {2025-12-19},
	journal = {Journal of the American Statistical Association},
	author = {Gneiting, Tilmann and Raftery, Adrian E},
	month = mar,
	year = {2007},
	pages = {359--378},
}

@article{ferreira_optimal_2015,
	title = {Optimal {Design} in {Geostatistics} under {Preferential} {Sampling}},
	volume = {10},
	issn = {1936-0975, 1931-6690},
	url = {https://projecteuclid.org/journals/bayesian-analysis/volume-10/issue-3/Optimal-Design-in-Geostatistics-under-Preferential-Sampling/10.1214/15-BA944.full},
	doi = {10.1214/15-BA944},
	number = {3},
	urldate = {2025-12-01},
	journal = {Bayesian Analysis},
	author = {Ferreira, Gustavo da Silva and Gamerman, Dani},
	month = sep,
	year = {2015},
	note = {Publisher: International Society for Bayesian Analysis},
	pages = {711--735},
}

@article{wang_prediction_2020,
	title = {On {Prediction} {Properties} of {Kriging}: {Uniform} {Error} {Bounds} and {Robustness}},
	volume = {115},
	issn = {0162-1459},
	shorttitle = {On {Prediction} {Properties} of {Kriging}},
	url = {https://doi.org/10.1080/01621459.2019.1598868},
	doi = {10.1080/01621459.2019.1598868},
	number = {530},
	urldate = {2025-12-01},
	journal = {Journal of the American Statistical Association},
	author = {Wang, Wenjia and Tuo, Rui and Jeff Wu, C. F.},
	month = apr,
	year = {2020},
	note = {Publisher: ASA Website
\_eprint: https://doi.org/10.1080/01621459.2019.1598868},
	pages = {920--930},
}

@article{katzfuss_general_2021,
	title = {A {General} {Framework} for {Vecchia} {Approximations} of {Gaussian} {Processes}},
	volume = {36},
	issn = {0883-4237, 2168-8745},
	url = {https://projecteuclid.org/journals/statistical-science/volume-36/issue-1/A-General-Framework-for-Vecchia-Approximations-of-Gaussian-Processes/10.1214/19-STS755.full},
	doi = {10.1214/19-STS755},
	number = {1},
	urldate = {2025-11-26},
	journal = {Statistical Science},
	author = {Katzfuss, Matthias and Guinness, Joseph},
	month = feb,
	year = {2021},
	note = {Publisher: Institute of Mathematical Statistics},
	pages = {124--141},
}

@article{baddeley_nonparametric_2012,
	title = {Nonparametric estimation of the dependence of a spatial point process on spatial covariates},
	volume = {5},
	issn = {19387989, 19387997},
	url = {https://link.intlpress.com/JDetail/1806634541366358018},
	doi = {10.4310/SII.2012.v5.n2.a7},
	language = {en},
	number = {2},
	urldate = {2025-11-25},
	journal = {Statistics and Its Interface},
	author = {Baddeley, Adrian and Chang, Ya-Mei and Song, Yong and Turner, Rolf},
	year = {2012},
	pages = {221--236},
}

@article{cronie_non-model-based_2018,
	title = {A non-model-based approach to bandwidth selection for kernel estimators of spatial intensity functions},
	volume = {105},
	issn = {0006-3444},
	url = {https://doi.org/10.1093/biomet/asy001},
	doi = {10.1093/biomet/asy001},
	number = {2},
	urldate = {2025-11-21},
	journal = {Biometrika},
	author = {Cronie, O and Van Lieshout, M N M},
	month = jun,
	year = {2018},
	pages = {455--462},
}

@misc{biscio_nonparametric_2025,
	title = {Nonparametric intensity estimation of spatial point processes by random forests},
	url = {http://arxiv.org/abs/2511.09307},
	doi = {10.48550/arXiv.2511.09307},
	urldate = {2025-11-20},
	publisher = {arXiv},
	author = {Biscio, Christophe and Lavancier, Frédéric},
	month = nov,
	year = {2025},
	note = {arXiv:2511.09307 [stat]
version: 1},
}

@article{gilbert_consistency_2025,
	title = {Consistency of common spatial estimators under spatial confounding},
	volume = {112},
	issn = {1464-3510},
	url = {https://doi.org/10.1093/biomet/asae070},
	doi = {10.1093/biomet/asae070},
	number = {2},
	urldate = {2025-07-22},
	journal = {Biometrika},
	author = {Gilbert, Brian and Ogburn, Elizabeth L and Datta, Abhirup},
	month = apr,
	year = {2025},
	pages = {asae070},
}

@article{moreira_analysis_2022,
	title = {Analysis of presence-only data via exact {Bayes}, with model and effects identification},
	volume = {16},
	issn = {1932-6157},
	url = {https://projecteuclid.org/journals/annals-of-applied-statistics/volume-16/issue-3/Analysis-of-presence-only-data-via-exact-Bayes-with-model/10.1214/21-AOAS1569.full},
	doi = {10.1214/21-aoas1569},
	language = {en},
	number = {3},
	urldate = {2025-07-17},
	journal = {The Annals of Applied Statistics},
	author = {Moreira, Guido A. and Gamerman, Dani},
	month = sep,
	year = {2022},
	note = {Publisher: Institute of Mathematical Statistics},
}

@article{moreira_presence-only_2024,
	title = {Presence-{Only} for {Marked} {Point} {Process} {Under} {Preferential} {Sampling}},
	volume = {29},
	copyright = {2023 The Author(s)},
	issn = {1537-2693},
	url = {https://link.springer.com/article/10.1007/s13253-023-00558-x},
	doi = {10.1007/s13253-023-00558-x},
	language = {en},
	number = {1},
	urldate = {2025-07-17},
	journal = {Journal of Agricultural, Biological and Environmental Statistics},
	author = {Moreira, Guido A. and Menezes, Raquel and Wise, Laura},
	month = mar,
	year = {2024},
	note = {Company: Springer
Distributor: Springer
Institution: Springer
Label: Springer
Number: 1
Publisher: Springer US},
	pages = {92--109},
}

@phdthesis{yu_parametric_2022,
	title = {Parametric {Estimation} in {Spatial} {Regression} {Models}},
	language = {en},
	school = {University of Maryland, College Park},
	author = {Yu, Nathan},
	year = {2022},
}

@article{mardia_maximum_1984,
	title = {Maximum {Likelihood} {Estimation} of {Models} for {Residual} {Covariance} in {Spatial} {Regression}},
	volume = {71},
	issn = {0006-3444},
	url = {https://www.jstor.org/stable/2336405},
	doi = {10.2307/2336405},
	number = {1},
	urldate = {2025-01-31},
	journal = {Biometrika},
	author = {Mardia, K. V. and Marshall, R. J.},
	year = {1984},
	note = {Publisher: [Oxford University Press, Biometrika Trust]},
	pages = {135--146},
}

@article{zhang_fixed-domain_2024,
	title = {Fixed-{Domain} {Asymptotics} {Under} {Vecchia}’s {Approximation} of {Spatial} {Process} {Likelihoods}},
	volume = {34},
	issn = {1017-0405},
	url = {https://www.ncbi.nlm.nih.gov/pmc/articles/PMC11444644/},
	doi = {10.5705/ss.202021.0428},
	number = {4},
	urldate = {2024-12-03},
	journal = {Statistica Sinica},
	author = {Zhang, Lu and Tang, Wenpin and Banerjee, Sudipto},
	month = oct,
	year = {2024},
	pmid = {39355373},
	pmcid = {PMC11444644},
	pages = {1863--1881},
}

@article{stein_asymptotically_1988,
	title = {Asymptotically {Efficient} {Prediction} of a {Random} {Field} with a {Misspecified} {Covariance} {Function}},
	volume = {16},
	issn = {0090-5364, 2168-8966},
	url = {https://projecteuclid.org/journals/annals-of-statistics/volume-16/issue-1/Asymptotically-Efficient-Prediction-of-a-Random-Field-with-a-Misspecified/10.1214/aos/1176350690.full},
	doi = {10.1214/aos/1176350690},
	number = {1},
	urldate = {2024-11-20},
	journal = {The Annals of Statistics},
	author = {Stein, Michael L.},
	month = mar,
	year = {1988},
	note = {Publisher: Institute of Mathematical Statistics},
	pages = {55--63},
}

@article{kaufman_role_2013,
	title = {The role of the range parameter for estimation and prediction in geostatistics},
	volume = {100},
	issn = {0006-3444, 1464-3510},
	url = {https://academic.oup.com/biomet/article-lookup/doi/10.1093/biomet/ass079},
	doi = {10.1093/biomet/ass079},
	language = {en},
	number = {2},
	urldate = {2024-11-20},
	journal = {Biometrika},
	author = {Kaufman, C. G. and Shaby, B. A.},
	month = jun,
	year = {2013},
	pages = {473--484},
}

@article{tang_identifiability_2021,
	title = {On {Identifiability} and {Consistency} of {The} {Nugget} in {Gaussian} {Spatial} {Process} {Models}},
	volume = {83},
	copyright = {https://academic.oup.com/journals/pages/open\_access/funder\_policies/chorus/standard\_publication\_model},
	issn = {1369-7412, 1467-9868},
	url = {https://academic.oup.com/jrsssb/article/83/5/1044/7056082},
	doi = {10.1111/rssb.12472},
	language = {en},
	number = {5},
	urldate = {2024-11-20},
	journal = {Journal of the Royal Statistical Society Series B: Statistical Methodology},
	author = {Tang, Wenpin and Zhang, Lu and Banerjee, Sudipto},
	month = nov,
	year = {2021},
	pages = {1044--1070},
}

@article{stein_simple_1993,
	title = {A simple condition for asymptotic optimality of linear predictions of random fields},
	volume = {17},
	issn = {0167-7152},
	url = {https://www.sciencedirect.com/science/article/pii/016771529390261G},
	doi = {10.1016/0167-7152(93)90261-G},
	number = {5},
	urldate = {2024-06-12},
	journal = {Statistics \& Probability Letters},
	author = {Stein, Michael L.},
	month = aug,
	year = {1993},
	pages = {399--404},
}

@article{loh_fixed-domain_2021,
	title = {On fixed-domain asymptotics, parameter estimation and isotropic {Gaussian} random fields with {Matérn} covariance functions},
	volume = {49},
	issn = {0090-5364},
	url = {https://projecteuclid.org/journals/annals-of-statistics/volume-49/issue-6/On-fixed-domain-asymptotics-parameter-estimation-and-isotropic-Gaussian-random/10.1214/21-AOS2077.full},
	doi = {10.1214/21-AOS2077},
	language = {en},
	number = {6},
	urldate = {2023-10-24},
	journal = {The Annals of Statistics},
	author = {Loh, Wei-Liem and Sun, Saifei and Wen, Jun},
	month = dec,
	year = {2021},
}

@article{stein_uniform_1990,
	title = {Uniform {Asymptotic} {Optimality} of {Linear} {Predictions} of a {Random} {Field} {Using} an {Incorrect} {Second}-{Order} {Structure}},
	volume = {18},
	issn = {0090-5364, 2168-8966},
	url = {https://projecteuclid.org/journals/annals-of-statistics/volume-18/issue-2/Uniform-Asymptotic-Optimality-of-Linear-Predictions-of-a-Random-Field/10.1214/aos/1176347629.full},
	doi = {10.1214/aos/1176347629},
	number = {2},
	urldate = {2023-10-24},
	journal = {The Annals of Statistics},
	author = {Stein, Michael},
	month = jun,
	year = {1990},
	note = {Publisher: Institute of Mathematical Statistics},
	pages = {850--872},
}

@article{putter_effect_2001,
	title = {On the {Effect} of {Covariance} {Function} {Estimation} on the {Accuracy} of {Kriging} {Predictors}},
	volume = {7},
	issn = {1350-7265},
	url = {https://www.jstor.org/stable/3318494},
	doi = {10.2307/3318494},
	number = {3},
	urldate = {2023-10-24},
	journal = {Bernoulli},
	author = {Putter, Hein and Young, G. Alastair},
	year = {2001},
	note = {Publisher: International Statistical Institute (ISI) and Bernoulli Society for Mathematical Statistics and Probability},
	pages = {421--438},
}

@article{bevilacqua_comparing_2015,
	title = {Comparing composite likelihood methods based on pairs for spatial {Gaussian} random fields},
	volume = {25},
	issn = {1573-1375},
	url = {https://doi.org/10.1007/s11222-014-9460-6},
	doi = {10.1007/s11222-014-9460-6},
	language = {en},
	number = {5},
	urldate = {2023-02-21},
	journal = {Statistics and Computing},
	author = {Bevilacqua, Moreno and Gaetan, Carlo},
	month = sep,
	year = {2015},
	pages = {877--892},
}

@article{bachoc_asymptotic_2014,
	title = {Asymptotic analysis of the role of spatial sampling for covariance parameter estimation of {Gaussian} processes},
	volume = {125},
	issn = {0047-259X},
	url = {https://www.sciencedirect.com/science/article/pii/S0047259X13002571},
	doi = {10.1016/j.jmva.2013.11.015},
	urldate = {2023-08-21},
	journal = {Journal of Multivariate Analysis},
	author = {Bachoc, François},
	month = mar,
	year = {2014},
	pages = {1--35},
}

@article{bachoc_composite_2019,
	title = {Composite likelihood estimation for a {Gaussian} process under fixed domain asymptotics},
	volume = {174},
	issn = {0047-259X},
	url = {https://www.sciencedirect.com/science/article/pii/S0047259X18303841},
	doi = {10.1016/j.jmva.2019.104534},
	language = {en},
	urldate = {2023-02-21},
	journal = {Journal of Multivariate Analysis},
	author = {Bachoc, François and Bevilacqua, Moreno and Velandia, Daira},
	month = nov,
	year = {2019},
	pages = {104534},
}

@article{silva_exact_2024,
	title = {Exact {Bayesian} {Geostatistics} {Under} {Preferential} {Sampling}},
	volume = {-1},
	issn = {1936-0975, 1931-6690},
	url = {https://projecteuclid.org/journals/bayesian-analysis/advance-publication/Exact-Bayesian-Geostatistics-Under-Preferential-Sampling/10.1214/24-BA1460.full},
	doi = {10.1214/24-BA1460},
	number = {-1},
	urldate = {2025-05-22},
	journal = {Bayesian Analysis},
	author = {Silva, Douglas Mateus da and Gamerman, Dani},
	month = jan,
	year = {2024},
	note = {Publisher: International Society for Bayesian Analysis},
	pages = {1--29},
}

@article{zhang_inconsistent_2004,
	title = {Inconsistent {Estimation} and {Asymptotically} {Equal} {Interpolations} in {Model}-{Based} {Geostatistics}},
	volume = {99},
	issn = {0162-1459, 1537-274X},
	url = {http://www.tandfonline.com/doi/abs/10.1198/016214504000000241},
	doi = {10.1198/016214504000000241},
	language = {en},
	number = {465},
	urldate = {2025-03-25},
	journal = {Journal of the American Statistical Association},
	author = {Zhang, Hao},
	month = mar,
	year = {2004},
	pages = {250--261},
}

@article{cecconi_preferential_2016,
	title = {Preferential sampling in veterinary parasitological surveillance},
	volume = {11},
	copyright = {Copyright (c) 2016 Lorenzo Cecconi, Annibale Biggeri, Laura Grisotto, Veronica Berrocal, Laura Rinaldi, Vincenzo Musella, Giuseppe Cringoli, Dolores Catelan},
	issn = {1970-7096},
	url = {https://www.geospatialhealth.net/gh/article/view/412},
	doi = {10.4081/gh.2016.412},
	language = {en},
	number = {1},
	urldate = {2025-03-07},
	journal = {Geospatial Health},
	author = {Cecconi, Lorenzo and Biggeri, Annibale and Grisotto, Laura and Berrocal, Veronica and Rinaldi, Laura and Musella, Vincenzo and Cringoli, Giuseppe and Catelan, Dolores},
	month = apr,
	year = {2016},
	note = {Number: 1},
}

@misc{bachoc_asymptotic_2020,
	title = {Asymptotic analysis of maximum likelihood estimation of covariance parameters for {Gaussian} processes: an introduction with proofs},
	shorttitle = {Asymptotic analysis of maximum likelihood estimation of covariance parameters for {Gaussian} processes},
	url = {http://arxiv.org/abs/2009.07002},
	doi = {10.48550/arXiv.2009.07002},
	urldate = {2025-03-07},
	publisher = {arXiv},
	author = {Bachoc, François},
	month = sep,
	year = {2020},
	note = {arXiv:2009.07002 [math]},
}

@article{lee_constructing_2011,
	title = {Constructing representative air quality indicators with measures of uncertainty},
	volume = {174},
	copyright = {© 2011 Royal Statistical Society},
	issn = {1467-985X},
	url = {https://rss.onlinelibrary.wiley.com/doi/abs/10.1111/j.1467-985X.2010.00658.x},
	doi = {10.1111/j.1467-985X.2010.00658.x},
	language = {en},
	number = {1},
	urldate = {2020-08-11},
	journal = {Journal of the Royal Statistical Society: Series A (Statistics in Society)},
	author = {Lee, Duncan and Ferguson, Claire and Scott, E. Marian},
	year = {2011},
	note = {\_eprint: https://rss.onlinelibrary.wiley.com/doi/pdf/10.1111/j.1467-985X.2010.00658.x},
	pages = {109--126},
}

@article{shirota_preferential_2022,
	title = {Preferential sampling for bivariate spatial data},
	volume = {51},
	issn = {2211-6753},
	url = {https://www.sciencedirect.com/science/article/pii/S2211675322000458},
	doi = {10.1016/j.spasta.2022.100674},
	urldate = {2024-05-09},
	journal = {Spatial Statistics},
	author = {Shirota, Shinichiro and Gelfand, Alan E.},
	month = oct,
	year = {2022},
	pages = {100674},
}

@article{rue_approximate_2009,
	title = {Approximate {Bayesian} inference for latent {Gaussian} models by using integrated nested {Laplace} approximations},
	volume = {71},
	copyright = {© 2009 Royal Statistical Society},
	issn = {1467-9868},
	url = {https://onlinelibrary.wiley.com/doi/abs/10.1111/j.1467-9868.2008.00700.x},
	doi = {10.1111/j.1467-9868.2008.00700.x},
	language = {en},
	number = {2},
	urldate = {2023-12-11},
	journal = {Journal of the Royal Statistical Society: Series B (Statistical Methodology)},
	author = {Rue, Håvard and Martino, Sara and Chopin, Nicolas},
	year = {2009},
	note = {\_eprint: https://rss.onlinelibrary.wiley.com/doi/pdf/10.1111/j.1467-9868.2008.00700.x},
	pages = {319--392},
}

@article{manceur_inferring_2014,
	title = {Inferring model-based probability of occurrence from preferentially sampled data with uncertain absences using expert knowledge},
	volume = {5},
	copyright = {© 2014 The Authors. Methods in Ecology and Evolution © 2014 British Ecological Society},
	issn = {2041-210X},
	url = {https://onlinelibrary.wiley.com/doi/abs/10.1111/2041-210X.12224},
	doi = {10.1111/2041-210X.12224},
	language = {en},
	number = {8},
	urldate = {2024-12-07},
	journal = {Methods in Ecology and Evolution},
	author = {Manceur, Ameur M. and Kühn, Ingolf},
	year = {2014},
	note = {\_eprint: https://onlinelibrary.wiley.com/doi/pdf/10.1111/2041-210X.12224},
	pages = {739--750},
}

@article{van_lieshout_infill_2021,
	title = {Infill asymptotics for adaptive kernel estimators of spatial intensity},
	volume = {63},
	copyright = {© 2021 John Wiley \& Sons Australia, Ltd},
	issn = {1467-842X},
	url = {https://onlinelibrary.wiley.com/doi/abs/10.1111/anzs.12319},
	doi = {10.1111/anzs.12319},
	language = {en},
	number = {1},
	urldate = {2025-02-25},
	journal = {Australian \& New Zealand Journal of Statistics},
	author = {van Lieshout, M.n.m.},
	year = {2021},
	note = {\_eprint: https://onlinelibrary.wiley.com/doi/pdf/10.1111/anzs.12319},
	pages = {159--181},
}

@article{varin_overview_2011,
	title = {An {Overview} of {Composite} {Likelihood} {Methods}},
	volume = {21},
	issn = {1017-0405},
	url = {https://www.jstor.org/stable/24309261},
	number = {1},
	urldate = {2025-02-21},
	journal = {Statistica Sinica},
	author = {Varin, Cristiano and Reid, Nancy and Firth, David},
	year = {2011},
	note = {Publisher: Institute of Statistical Science, Academia Sinica},
	pages = {5--42},
}

@book{cressie_statistics_2015,
	title = {Statistics for {Spatio}-{Temporal} {Data}},
	isbn = {978-1-119-24304-5},
	language = {en},
	publisher = {John Wiley \& Sons},
	author = {Cressie, Noel and Wikle, Christopher K.},
	month = nov,
	year = {2015},
	note = {Google-Books-ID: 4L\_dCgAAQBAJ},
}

@misc{bolin_spatial_2024,
	title = {Spatial confounding under infill asymptotics},
	url = {http://arxiv.org/abs/2403.18961},
	doi = {10.48550/arXiv.2403.18961},
	urldate = {2025-02-07},
	publisher = {arXiv},
	author = {Bolin, David and Wallin, Jonas},
	month = mar,
	year = {2024},
	note = {arXiv:2403.18961 [math]},
}

@article{stein_bounds_1990,
	title = {Bounds on the {Efficiency} of {Linear} {Predictions} {Using} an {Incorrect} {Covariance} {Function}},
	volume = {18},
	issn = {0090-5364, 2168-8966},
	url = {https://projecteuclid.org/journals/annals-of-statistics/volume-18/issue-3/Bounds-on-the-Efficiency-of-Linear-Predictions-Using-an-Incorrect/10.1214/aos/1176347742.full},
	doi = {10.1214/aos/1176347742},
	number = {3},
	urldate = {2024-11-20},
	journal = {The Annals of Statistics},
	author = {Stein, Michael L.},
	month = sep,
	year = {1990},
	note = {Publisher: Institute of Mathematical Statistics},
	pages = {1116--1138},
}

@article{guinness_permutation_2018,
	title = {Permutation and {Grouping} {Methods} for {Sharpening} {Gaussian} {Process} {Approximations}},
	volume = {60},
	issn = {0040-1706},
	url = {https://doi.org/10.1080/00401706.2018.1437476},
	doi = {10.1080/00401706.2018.1437476},
	number = {4},
	urldate = {2024-06-10},
	journal = {Technometrics},
	author = {Guinness, Joseph},
	month = oct,
	year = {2018},
	pmid = {31447491},
	note = {Publisher: Taylor \& Francis
\_eprint: https://doi.org/10.1080/00401706.2018.1437476},
	pages = {415--429},
}

@article{berman_estimating_1989,
	title = {Estimating {Weighted} {Integrals} of the {Second}-{Order} {Intensity} of a {Spatial} {Point} {Process}},
	volume = {51},
	issn = {0035-9246},
	url = {https://www.jstor.org/stable/2345843},
	number = {1},
	urldate = {2024-05-23},
	journal = {Journal of the Royal Statistical Society. Series B (Methodological)},
	author = {Berman, Mark and Diggle, Peter},
	year = {1989},
	note = {Publisher: [Royal Statistical Society, Wiley]},
	pages = {81--92},
}

@article{lindgren_explicit_2011,
	title = {An explicit link between {Gaussian} fields and {Gaussian} {Markov} random fields: the stochastic partial differential equation approach},
	volume = {73},
	copyright = {© 2011 Royal Statistical Society},
	issn = {1467-9868},
	shorttitle = {An explicit link between {Gaussian} fields and {Gaussian} {Markov} random fields},
	url = {https://onlinelibrary.wiley.com/doi/abs/10.1111/j.1467-9868.2011.00777.x},
	doi = {10.1111/j.1467-9868.2011.00777.x},
	language = {en},
	number = {4},
	urldate = {2024-05-20},
	journal = {Journal of the Royal Statistical Society: Series B (Statistical Methodology)},
	author = {Lindgren, Finn and Rue, Håvard and Lindström, Johan},
	year = {2011},
	note = {\_eprint: https://onlinelibrary.wiley.com/doi/pdf/10.1111/j.1467-9868.2011.00777.x},
	pages = {423--498},
}

@article{fandos_dynamic_2021,
	title = {Dynamic multistate occupancy modeling to evaluate population dynamics under a scenario of preferential sampling},
	volume = {12},
	copyright = {© 2021 The Authors.},
	issn = {2150-8925},
	url = {https://onlinelibrary.wiley.com/doi/abs/10.1002/ecs2.3469},
	doi = {10.1002/ecs2.3469},
	language = {en},
	number = {4},
	urldate = {2024-05-18},
	journal = {Ecosphere},
	author = {Fandos, Guillermo and Kéry, Marc and Cano-Alonso, Luis Santiago and Carbonell, Isidoro and Luis Tellería, José},
	year = {2021},
	note = {\_eprint: https://onlinelibrary.wiley.com/doi/pdf/10.1002/ecs2.3469},
	pages = {e03469},
}

@article{kristensen_tmb_2016,
	title = {{TMB}: {Automatic} {Differentiation} and {Laplace} {Approximation}},
	volume = {70},
	copyright = {Copyright (c) 2016 Kasper Kristensen, Anders Nielsen, Casper W. Berg, Hans Skaug, Bradley M. Bell},
	issn = {1548-7660},
	shorttitle = {{TMB}},
	url = {https://doi.org/10.18637/jss.v070.i05},
	doi = {10.18637/jss.v070.i05},
	language = {en},
	urldate = {2024-05-18},
	journal = {Journal of Statistical Software},
	author = {Kristensen, Kasper and Nielsen, Anders and Berg, Casper W. and Skaug, Hans and Bell, Bradley M.},
	month = apr,
	year = {2016},
	pages = {1--21},
}

@article{ho_modelling_2008,
	title = {Modelling marked point patterns by intensity-marked {Cox} processes},
	volume = {78},
	issn = {0167-7152},
	url = {https://www.sciencedirect.com/science/article/pii/S0167715207003926},
	doi = {10.1016/j.spl.2007.11.013},
	language = {en},
	number = {10},
	urldate = {2022-09-14},
	journal = {Statistics \& Probability Letters},
	author = {Ho, Lai Ping and Stoyan, D.},
	month = aug,
	year = {2008},
	pages = {1194--1199},
}

@article{conn_confronting_2017,
	title = {Confronting preferential sampling when analysing population distributions: diagnosis and model-based triage},
	volume = {8},
	copyright = {Published 2017. This article is a U.S. Government work and is in the public domain in the USA.},
	issn = {2041-210X},
	shorttitle = {Confronting preferential sampling when analysing population distributions},
	url = {https://besjournals.onlinelibrary.wiley.com/doi/abs/10.1111/2041-210X.12803},
	doi = {10.1111/2041-210X.12803},
	language = {en},
	number = {11},
	urldate = {2020-08-11},
	journal = {Methods in Ecology and Evolution},
	author = {Conn, Paul B. and Thorson, James T. and Johnson, Devin S.},
	year = {2017},
	note = {\_eprint: https://besjournals.onlinelibrary.wiley.com/doi/pdf/10.1111/2041-210X.12803},
	pages = {1535--1546},
}

@article{gelfand_effect_2012,
	title = {On the effect of preferential sampling in spatial prediction},
	volume = {23},
	copyright = {Copyright © 2012 John Wiley \& Sons, Ltd.},
	issn = {1099-095X},
	url = {https://onlinelibrary.wiley.com/doi/abs/10.1002/env.2169},
	doi = {10.1002/env.2169},
	language = {en},
	number = {7},
	urldate = {2020-08-11},
	journal = {Environmetrics},
	author = {Gelfand, Alan E. and Sahu, Sujit K. and Holland, David M.},
	year = {2012},
	note = {\_eprint: https://onlinelibrary.wiley.com/doi/pdf/10.1002/env.2169},
	pages = {565--578},
}

@article{gray_design_2023,
	title = {A design utility approach for preferentially sampled spatial data},
	volume = {72},
	issn = {0035-9254},
	url = {https://doi.org/10.1093/jrsssc/qlad040},
	doi = {10.1093/jrsssc/qlad040},
	number = {4},
	urldate = {2024-04-24},
	journal = {Journal of the Royal Statistical Society Series C: Applied Statistics},
	author = {Gray, Elizabeth J and Evangelou, Evangelos},
	month = aug,
	year = {2023},
	pages = {1041--1063},
}

@incollection{mateu_accounting_2012,
	address = {Chichester, UK},
	title = {Accounting for {Design} in the {Analysis} of {Spatial} {Data}},
	isbn = {978-1-118-44186-2 978-0-470-97429-2},
	url = {https://onlinelibrary.wiley.com/doi/10.1002/9781118441862.ch6},
	language = {en},
	urldate = {2022-08-04},
	booktitle = {Spatio-{Temporal} {Design}},
	publisher = {John Wiley \& Sons, Ltd},
	author = {Reich, Brian J. and Fuentes, Montserrat},
	editor = {Mateu, Jorge and Müller, Werner G.},
	month = oct,
	year = {2012},
	doi = {10.1002/9781118441862.ch6},
	pages = {131--141},
}

@article{schlather_detecting_2004,
	title = {Detecting dependence between marks and locations of marked point processes},
	volume = {66},
	issn = {1467-9868},
	url = {https://onlinelibrary.wiley.com/doi/abs/10.1046/j.1369-7412.2003.05343.x},
	doi = {10.1046/j.1369-7412.2003.05343.x},
	language = {en},
	number = {1},
	urldate = {2022-05-24},
	journal = {Journal of the Royal Statistical Society: Series B (Statistical Methodology)},
	author = {Schlather, Martin and Ribeiro Jr, Paulo J. and Diggle, Peter J.},
	year = {2004},
	note = {\_eprint: https://onlinelibrary.wiley.com/doi/pdf/10.1046/j.1369-7412.2003.05343.x},
	pages = {79--93},
}

@article{watson_perceptron_2021,
	title = {A perceptron for detecting the preferential sampling of locations and times chosen to monitor a spatio-temporal process},
	volume = {43},
	issn = {2211-6753},
	url = {https://www.sciencedirect.com/science/article/pii/S2211675321000105},
	doi = {10.1016/j.spasta.2021.100500},
	language = {en},
	urldate = {2022-09-13},
	journal = {Spatial Statistics},
	author = {Watson, Joe},
	month = jun,
	year = {2021},
	pages = {100500},
}

@article{zidek_reducing_2014,
	title = {Reducing estimation bias in adaptively changing monitoring networks with preferential site selection},
	volume = {8},
	issn = {1932-6157, 1941-7330},
	url = {https://projecteuclid.org/journals/annals-of-applied-statistics/volume-8/issue-3/Reducing-estimation-bias-in-adaptively-changing-monitoring-networks-with-preferential/10.1214/14-AOAS745.full},
	doi = {10.1214/14-AOAS745},
	number = {3},
	urldate = {2021-10-28},
	journal = {The Annals of Applied Statistics},
	author = {Zidek, James V. and Shaddick, Gavin and Taylor, Carolyn G.},
	month = sep,
	year = {2014},
	note = {Publisher: Institute of Mathematical Statistics},
	pages = {1640--1670},
}

@article{watson_general_2019,
	title = {A general theory for preferential sampling in environmental networks},
	volume = {13},
	issn = {1932-6157, 1941-7330},
	url = {https://projecteuclid.org/journals/annals-of-applied-statistics/volume-13/issue-4/A-general-theory-for-preferential-sampling-in-environmental-networks/10.1214/19-AOAS1288.full},
	doi = {10.1214/19-AOAS1288},
	number = {4},
	urldate = {2021-12-13},
	journal = {The Annals of Applied Statistics},
	author = {Watson, Joe and Zidek, James V. and Shaddick, Gavin},
	month = dec,
	year = {2019},
	note = {Publisher: Institute of Mathematical Statistics},
	pages = {2662--2700},
}

@article{amaral_model-based_2023,
	title = {Model-{Based} {Geostatistics} {Under} {Spatially} {Varying} {Preferential} {Sampling}},
	issn = {1537-2693},
	url = {https://doi.org/10.1007/s13253-023-00571-0},
	doi = {10.1007/s13253-023-00571-0},
	language = {en},
	urldate = {2023-10-01},
	journal = {Journal of Agricultural, Biological and Environmental Statistics},
	author = {Amaral, André Victor Ribeiro and Krainski, Elias Teixeira and Zhong, Ruiman and Moraga, Paula},
	month = sep,
	year = {2023},
}

@article{dinsdale_methods_2019,
	title = {Methods for preferential sampling in geostatistics},
	volume = {68},
	issn = {1467-9876},
	url = {https://onlinelibrary.wiley.com/doi/abs/10.1111/rssc.12286},
	doi = {10.1111/rssc.12286},
	language = {en},
	number = {1},
	urldate = {2021-10-28},
	journal = {Journal of the Royal Statistical Society: Series C (Applied Statistics)},
	author = {Dinsdale, Daniel and Salibian-Barrera, Matias},
	year = {2019},
	note = {\_eprint: https://onlinelibrary.wiley.com/doi/pdf/10.1111/rssc.12286},
	pages = {181--198},
}

@article{diggle_geostatistical_2010,
	title = {Geostatistical inference under preferential sampling},
	volume = {59},
	issn = {1467-9876},
	url = {https://onlinelibrary.wiley.com/doi/abs/10.1111/j.1467-9876.2009.00701.x},
	doi = {10.1111/j.1467-9876.2009.00701.x},
	language = {en},
	number = {2},
	urldate = {2023-02-09},
	journal = {Journal of the Royal Statistical Society: Series C (Applied Statistics)},
	author = {Diggle, Peter J. and Menezes, Raquel and Su, Ting-li},
	year = {2010},
	note = {\_eprint: https://rss.onlinelibrary.wiley.com/doi/pdf/10.1111/j.1467-9876.2009.00701.x},
	pages = {191--232},
}

@article{diggle_model-based_1998,
	title = {Model-based geostatistics},
	volume = {47},
	issn = {1467-9876},
	url = {https://rss.onlinelibrary.wiley.com/doi/abs/10.1111/1467-9876.00113},
	doi = {10.1111/1467-9876.00113},
	language = {en},
	number = {3},
	urldate = {2020-07-26},
	journal = {Journal of the Royal Statistical Society: Series C (Applied Statistics)},
	author = {Diggle, P. J. and Tawn, J. A. and Moyeed, R. A.},
	year = {1998},
	note = {\_eprint: https://rss.onlinelibrary.wiley.com/doi/pdf/10.1111/1467-9876.00113},
	pages = {299--350},
}

@article{vedensky_look_2023,
	title = {A {Look} into the {Problem} of {Preferential} {Sampling} through the {Lens} of {Survey} {Statistics}},
	volume = {77},
	issn = {0003-1305},
	url = {https://doi.org/10.1080/00031305.2022.2143898},
	doi = {10.1080/00031305.2022.2143898},
	number = {3},
	urldate = {2024-02-20},
	journal = {The American Statistician},
	author = {Vedensky, Daniel and Parker, Paul A. and Holan, Scott H.},
	month = jul,
	year = {2023},
	note = {Publisher: Taylor \& Francis
\_eprint: https://doi.org/10.1080/00031305.2022.2143898},
	pages = {313--322},
}

@article{huser_vecchia_2023,
	title = {Vecchia {Likelihood} {Approximation} for {Accurate} and {Fast} {Inference} with {Intractable} {Spatial} {Max}-{Stable} {Models}},
	volume = {0},
	issn = {1061-8600},
	url = {https://doi.org/10.1080/10618600.2023.2285332},
	doi = {10.1080/10618600.2023.2285332},
	number = {0},
	urldate = {2024-02-08},
	journal = {Journal of Computational and Graphical Statistics},
	author = {Huser, Raphaël and Stein, Michael L. and Zhong, Peng},
	year = {2023},
	note = {Publisher: Taylor \& Francis
\_eprint: https://doi.org/10.1080/10618600.2023.2285332},
	pages = {1--22},
}

@incollection{scott_kernel_1992,
	title = {Kernel {Density} {Estimators}},
	copyright = {Copyright © 1992 John Wiley \& Sons, Inc.},
	isbn = {978-0-470-31684-9},
	url = {https://onlinelibrary.wiley.com/doi/abs/10.1002/9780470316849.ch6},
	language = {en},
	urldate = {2023-12-11},
	booktitle = {Multivariate {Density} {Estimation}},
	publisher = {John Wiley \& Sons, Ltd},
	author = {Scott, D.W.},
	year = {1992},
	doi = {10.1002/9780470316849.ch6},
	note = {Section: 6
\_eprint: https://onlinelibrary.wiley.com/doi/pdf/10.1002/9780470316849.ch6},
	pages = {125--193},
}

@incollection{loader_density_1999,
	address = {New York, NY},
	series = {Statistics and {Computing}},
	title = {Density {Estimation}},
	isbn = {978-0-387-22732-0},
	url = {https://doi.org/10.1007/0-387-22732-6_5},
	language = {en},
	urldate = {2023-12-11},
	booktitle = {Local {Regression} and {Likelihood}},
	publisher = {Springer},
	author = {Loader, Clive},
	editor = {Loader, Clive},
	year = {1999},
	doi = {10.1007/0-387-22732-6_5},
	pages = {79--100},
}

@article{diggle_kernel_1985,
	title = {A {Kernel} {Method} for {Smoothing} {Point} {Process} {Data}},
	volume = {34},
	copyright = {© 1985 Royal Statistical Society},
	issn = {1467-9876},
	url = {https://onlinelibrary.wiley.com/doi/abs/10.2307/2347366},
	doi = {10.2307/2347366},
	language = {en},
	number = {2},
	urldate = {2023-12-07},
	journal = {Journal of the Royal Statistical Society: Series C (Applied Statistics)},
	author = {Diggle, Peter},
	year = {1985},
	note = {\_eprint: https://onlinelibrary.wiley.com/doi/pdf/10.2307/2347366},
	pages = {138--147},
}

@article{guan_consistent_2008,
	title = {On {Consistent} {Nonparametric} {Intensity} {Estimation} for {Inhomogeneous} {Spatial} {Point} {Processes}},
	volume = {103},
	issn = {0162-1459, 1537-274X},
	url = {https://www.tandfonline.com/doi/full/10.1198/016214508000000526},
	doi = {10.1198/016214508000000526},
	language = {en},
	number = {483},
	urldate = {2023-12-07},
	journal = {Journal of the American Statistical Association},
	author = {Guan, Yongtao},
	month = sep,
	year = {2008},
	pages = {1238--1247},
}

@article{fithian_bias_2015,
	title = {Bias correction in species distribution models: pooling survey and collection data for multiple species},
	volume = {6},
	issn = {2041-210X},
	shorttitle = {Bias correction in species distribution models},
	url = {https://onlinelibrary.wiley.com/doi/abs/10.1111/2041-210X.12242},
	doi = {10.1111/2041-210X.12242},
	language = {en},
	number = {4},
	urldate = {2021-12-10},
	journal = {Methods in Ecology and Evolution},
	author = {Fithian, William and Elith, Jane and Hastie, Trevor and Keith, David A.},
	year = {2015},
	note = {\_eprint: https://onlinelibrary.wiley.com/doi/pdf/10.1111/2041-210X.12242},
	pages = {424--438},
}

@article{guinness_gaussian_2021,
	title = {Gaussian process learning via {Fisher} scoring of {Vecchia}’s approximation},
	volume = {31},
	issn = {1573-1375},
	url = {https://doi.org/10.1007/s11222-021-09999-1},
	doi = {10.1007/s11222-021-09999-1},
	language = {en},
	number = {3},
	urldate = {2023-10-26},
	journal = {Statistics and Computing},
	author = {Guinness, Joseph},
	month = mar,
	year = {2021},
	pages = {25},
}

@article{katzfuss_vecchia_2020,
	title = {Vecchia {Approximations} of {Gaussian}-{Process} {Predictions}},
	volume = {25},
	issn = {1537-2693},
	url = {https://doi.org/10.1007/s13253-020-00401-7},
	doi = {10.1007/s13253-020-00401-7},
	language = {en},
	number = {3},
	urldate = {2023-02-09},
	journal = {Journal of Agricultural, Biological and Environmental Statistics},
	author = {Katzfuss, Matthias and Guinness, Joseph and Gong, Wenlong and Zilber, Daniel},
	month = sep,
	year = {2020},
	pages = {383--414},
}

@article{simpson_going_2016,
	title = {Going off grid: computationally efficient inference for log-{Gaussian} {Cox} processes},
	volume = {103},
	issn = {0006-3444},
	shorttitle = {Going off grid},
	url = {https://doi.org/10.1093/biomet/asv064},
	doi = {10.1093/biomet/asv064},
	number = {1},
	urldate = {2022-01-27},
	journal = {Biometrika},
	author = {Simpson, D. and Illian, J. B. and Lindgren, F. and Sørbye, S. H. and Rue, H.},
	month = mar,
	year = {2016},
	pages = {49--70},
}

@article{pennino_accounting_2019,
	title = {Accounting for preferential sampling in species distribution models},
	volume = {9},
	copyright = {© 2018 The Authors. Ecology and Evolution  published by John Wiley \& Sons Ltd.},
	issn = {2045-7758},
	url = {https://onlinelibrary.wiley.com/doi/abs/10.1002/ece3.4789},
	doi = {10.1002/ece3.4789},
	language = {en},
	number = {1},
	urldate = {2020-08-11},
	journal = {Ecology and Evolution},
	author = {Pennino, Maria Grazia and Paradinas, Iosu and Illian, Janine B. and Muñoz, Facundo and Bellido, José María and López‐Quílez, Antonio and Conesa, David},
	year = {2019},
	note = {\_eprint: https://onlinelibrary.wiley.com/doi/pdf/10.1002/ece3.4789},
	pages = {653--663},
}

@article{pati_bayesian_2011,
	title = {Bayesian geostatistical modelling with informative sampling locations},
	volume = {98},
	issn = {0006-3444},
	url = {https://www.ncbi.nlm.nih.gov/pmc/articles/PMC3744635/},
	doi = {10.1093/biomet/asq067},
	number = {1},
	urldate = {2020-08-11},
	journal = {Biometrika},
	author = {Pati, D. and Reich, B. J. and Dunson, D. B.},
	month = mar,
	year = {2011},
	pmid = {23956461},
	pmcid = {PMC3744635},
	pages = {35--48},
}

@article{karcher_quantifying_2016,
	title = {Quantifying and {Mitigating} the {Effect} of {Preferential} {Sampling} on {Phylodynamic} {Inference}},
	volume = {12},
	issn = {1553-7358},
	url = {https://journals.plos.org/ploscompbiol/article?id=10.1371/journal.pcbi.1004789},
	doi = {10.1371/journal.pcbi.1004789},
	language = {en},
	number = {3},
	urldate = {2023-09-20},
	journal = {PLOS Computational Biology},
	author = {Karcher, Michael D. and Palacios, Julia A. and Bedford, Trevor and Suchard, Marc A. and Minin, Vladimir N.},
	month = mar,
	year = {2016},
	note = {Publisher: Public Library of Science},
	pages = {e1004789},
}

@article{gelfand_preferential_2019,
	title = {Preferential sampling for presence/absence data and for fusion of presence/absence data with presence-only data},
	volume = {89},
	issn = {1557-7015},
	url = {https://onlinelibrary.wiley.com/doi/abs/10.1002/ecm.1372},
	doi = {10.1002/ecm.1372},
	language = {en},
	number = {3},
	urldate = {2023-02-16},
	journal = {Ecological Monographs},
	author = {Gelfand, Alan E. and Shirota, Shinichiro},
	year = {2019},
	note = {\_eprint: https://onlinelibrary.wiley.com/doi/pdf/10.1002/ecm.1372},
	pages = {e01372},
}

@article{ferreira_geostatistics_2020,
	title = {Geostatistics under preferential sampling in the presence of local repulsion effects},
	volume = {27},
	issn = {1352-8505, 1573-3009},
	url = {https://link.springer.com/10.1007/s10651-020-00458-0},
	doi = {10.1007/s10651-020-00458-0},
	language = {en},
	number = {3},
	urldate = {2022-04-16},
	journal = {Environmental and Ecological Statistics},
	author = {Ferreira, Gustavo da Silva},
	month = sep,
	year = {2020},
	pages = {549--570},
}

@article{schliep_correcting_2023,
	title = {Correcting for informative sampling in spatial covariance estimation and kriging predictions},
	issn = {1435-5949},
	url = {https://doi.org/10.1007/s10109-023-00426-9},
	doi = {10.1007/s10109-023-00426-9},
	language = {en},
	urldate = {2023-09-28},
	journal = {Journal of Geographical Systems},
	author = {Schliep, Erin M. and Wikle, Christopher K. and Daw, Ranadeep},
	month = sep,
	year = {2023},
}

@article{conroy_shared_2023,
	title = {A {Shared} {Latent} {Process} {Model} to {Correct} for {Preferential} {Sampling} in {Disease} {Surveillance} {Systems}},
	volume = {28},
	issn = {1537-2693},
	url = {https://doi.org/10.1007/s13253-023-00535-4},
	doi = {10.1007/s13253-023-00535-4},
	language = {en},
	number = {3},
	urldate = {2023-09-15},
	journal = {Journal of Agricultural, Biological and Environmental Statistics},
	author = {Conroy, Brian and Waller, Lance A. and Buller, Ian D. and Hacker, Gregory M. and Tucker, James R. and Novak, Mark G.},
	month = sep,
	year = {2023},
	pages = {483--501},
}

@article{datta_hierarchical_2016,
	title = {Hierarchical {Nearest}-{Neighbor} {Gaussian} {Process} {Models} for {Large} {Geostatistical} {Datasets}},
	volume = {111},
	issn = {0162-1459},
	url = {https://doi.org/10.1080/01621459.2015.1044091},
	doi = {10.1080/01621459.2015.1044091},
	number = {514},
	urldate = {2023-01-27},
	journal = {Journal of the American Statistical Association},
	author = {Datta, Abhirup and Banerjee, Sudipto and Finley, Andrew O. and Gelfand, Alan E.},
	month = apr,
	year = {2016},
	pmid = {29720777},
	note = {Publisher: Taylor \& Francis
\_eprint: https://doi.org/10.1080/01621459.2015.1044091},
	pages = {800--812},
}

@article{vecchia_estimation_1988,
	title = {Estimation and {Model} {Identification} for {Continuous} {Spatial} {Processes}},
	volume = {50},
	issn = {0035-9246},
	url = {https://www.jstor.org/stable/2345768},
	number = {2},
	urldate = {2023-02-10},
	journal = {Journal of the Royal Statistical Society. Series B (Methodological)},
	author = {Vecchia, A. V.},
	year = {1988},
	note = {Publisher: [Royal Statistical Society, Wiley]},
	pages = {297--312},
}

@article{heaton_case_2019,
	title = {A {Case} {Study} {Competition} {Among} {Methods} for {Analyzing} {Large} {Spatial} {Data}},
	volume = {24},
	issn = {1537-2693},
	url = {https://doi.org/10.1007/s13253-018-00348-w},
	doi = {10.1007/s13253-018-00348-w},
	language = {en},
	number = {3},
	urldate = {2021-03-22},
	journal = {Journal of Agricultural, Biological and Environmental Statistics},
	author = {Heaton, Matthew J. and Datta, Abhirup and Finley, Andrew O. and Furrer, Reinhard and Guinness, Joseph and Guhaniyogi, Rajarshi and Gerber, Florian and Gramacy, Robert B. and Hammerling, Dorit and Katzfuss, Matthias and Lindgren, Finn and Nychka, Douglas W. and Sun, Furong and Zammit-Mangion, Andrew},
	month = sep,
	year = {2019},
	pages = {398--425},
}

@article{simpson_penalising_2017,
	title = {Penalising {Model} {Component} {Complexity}: {A} {Principled}, {Practical} {Approach} to {Constructing} {Priors}},
	volume = {32},
	issn = {0883-4237, 2168-8745},
	shorttitle = {Penalising {Model} {Component} {Complexity}},
	url = {http://projecteuclid.org/journals/statistical-science/volume-32/issue-1/Penalising-Model-Component-Complexity--A-Principled-Practical-Approach-to/10.1214/16-STS576.full},
	doi = {10.1214/16-STS576},
	number = {1},
	urldate = {2022-11-08},
	journal = {Statistical Science},
	author = {Simpson, Daniel and Rue, Håvard and Riebler, Andrea and Martins, Thiago G. and Sørbye, Sigrunn H.},
	month = feb,
	year = {2017},
	note = {Publisher: Institute of Mathematical Statistics},
	pages = {1--28},
}

@article{fuglstad_constructing_2019,
	title = {Constructing {Priors} that {Penalize} the {Complexity} of {Gaussian} {Random} {Fields}},
	volume = {114},
	issn = {0162-1459, 1537-274X},
	url = {https://www.tandfonline.com/doi/full/10.1080/01621459.2017.1415907},
	doi = {10.1080/01621459.2017.1415907},
	language = {en},
	number = {525},
	urldate = {2022-01-14},
	journal = {Journal of the American Statistical Association},
	author = {Fuglstad, Geir-Arne and Simpson, Daniel and Lindgren, Finn and Rue, Håvard},
	month = jan,
	year = {2019},
	pages = {445--452},
}

@article{rinaldi_sheep_2015,
	title = {Sheep and {Fasciola} hepatica in {Europe}: the {GLOWORM} experience},
	volume = {9},
	copyright = {Copyright (c) 2015 Laura Rinaldi, Annibale Biggeri, Vincenzo Musella, Theo de Waal, Hubertus Hertzberg, Fabien Mavrot, Paul R. Torgerson, Nikolaos Selemetas, Tom Coll, Antonio Bosco, Laura Grisotto, Giuseppe Cringoli, Dolores Catelan},
	issn = {1970-7096},
	shorttitle = {Sheep and {Fasciola} hepatica in {Europe}},
	url = {https://www.geospatialhealth.net/index.php/gh/article/view/353},
	doi = {10.4081/gh.2015.353},
	language = {en},
	number = {2},
	urldate = {2020-08-11},
	journal = {Geospatial Health},
	author = {Rinaldi, Laura and Biggeri, Annibale and Musella, Vincenzo and Waal, Theo de and Hertzberg, Hubertus and Mavrot, Fabien and Torgerson, Paul R. and Selemetas, Nikolaos and Coll, Tom and Bosco, Antonio and Grisotto, Laura and Cringoli, Giuseppe and Catelan, Dolores},
	month = mar,
	year = {2015},
	note = {Number: 2},
	pages = {309--317},
}

@article{lee_impact_2015,
	title = {Impact of preferential sampling on exposure prediction and health effect inference in the context of air pollution epidemiology},
	volume = {26},
	copyright = {Copyright © 2015 John Wiley \& Sons, Ltd.},
	issn = {1099-095X},
	url = {https://onlinelibrary.wiley.com/doi/abs/10.1002/env.2334},
	doi = {10.1002/env.2334},
	language = {en},
	number = {4},
	urldate = {2020-08-11},
	journal = {Environmetrics},
	author = {Lee, A. and Szpiro, A. and Kim, S. Y. and Sheppard, L.},
	year = {2015},
	note = {\_eprint: https://onlinelibrary.wiley.com/doi/pdf/10.1002/env.2334},
	pages = {255--267},
}

@article{paci_spatial_2020,
	title = {Spatial hedonic modelling adjusted for preferential sampling},
	volume = {183},
	copyright = {© 2019 Royal Statistical Society},
	issn = {1467-985X},
	url = {https://rss.onlinelibrary.wiley.com/doi/abs/10.1111/rssa.12489},
	doi = {10.1111/rssa.12489},
	language = {en},
	number = {1},
	urldate = {2020-08-11},
	journal = {Journal of the Royal Statistical Society: Series A (Statistics in Society)},
	author = {Paci, Lucia and Gelfand, Alan E. and Beamonte, {and} María Asunción and Gargallo, Pilar and Salvador, Manuel},
	year = {2020},
	note = {\_eprint: https://rss.onlinelibrary.wiley.com/doi/pdf/10.1111/rssa.12489},
	pages = {169--192},
}

\clearpage

\section*{Figures and Tables}

\begin{table}[!ht]
\centering
\caption{Mean CRPS (SE) across the $B=500$ simulations for the synthetic data experiment with $N=100$ points. Bold indicates the best (lowest) value among all methods for the simulation scenario defined in each column.}
\label{tab:synthetic_crps}
\resizebox{\textwidth}{!}{ % Ensures table width fits within margins
\begin{tabular}{ll cccc cc}
\toprule
\textbf{Point process}& \textbf{Method} & \multicolumn{2}{c}{$\boldsymbol{\beta} = -1$} & \multicolumn{2}{c}{$\boldsymbol{\beta} = 1$} & \multicolumn{2}{c}{$\boldsymbol{\beta} = 2$} \\
\cmidrule(lr){3-4} \cmidrule(lr){5-6} \cmidrule(lr){7-8}
& & $\phi = 0.02$ & $\phi = 0.15$ & $\phi = 0.02$ & $\phi = 0.15$ & $\phi = 0.02$ & $\phi = 0.15$ \\
\midrule
\multirow{5}{*}{\shortstack[l]{LGCP}} 
& MLE & 1.035 (0.09) & 0.408 (0.09) & 1.030 (0.08) & 0.409 (0.08) & 1.789 (0.15) & 0.832 (0.27) \\
& INLA-SLP & \textbf{0.687 (0.10)} & \textbf{0.296 (0.04)} & \textbf{0.681 (0.06)} & \textbf{0.296 (0.04)} & \textbf{0.829 (0.13)} & \textbf{0.380 (0.08)} \\
& ISIW Known & 0.690 (0.10) & 0.352 (0.07) & 0.692 (0.10) & 0.352 (0.07) & 1.086 (0.39) & 0.613 (0.26) \\
& ISIW KIE & 0.975 (0.15) & 0.386 (0.09) & 0.986 (0.17) & 0.403 (0.18) & 1.689 (0.33) & 0.762 (0.30) \\
& ISIW KIE COV & 1.101 (0.14) & 0.390 (0.09) & 1.123 (0.16) & 0.390 (0.08) & 2.052 (0.34) & 0.806 (0.33) \\
\midrule
\multirow{5}{*}{\shortstack[l]{Thomas}} 
& MLE & 0.984 (0.09) & 0.435 (0.11) & 0.981 (0.08) & 0.435 (0.11) & 1.494 (0.24) & 0.840 (0.40) \\
& INLA-SLP & 0.809 (0.31) & 0.419 (0.18) & 0.728 (0.10) & 0.407 (0.14) & 1.804 (1.21) & 0.786 (0.58) \\
& ISIW Known & \textbf{0.688 (0.07)} & \textbf{0.384 (0.08)} & \textbf{0.698 (0.10)} & \textbf{0.384 (0.08)} & \textbf{0.959 (0.31)} & \textbf{0.674 (0.33)} \\
& ISIW KIE & 0.913 (0.15) & 0.416 (0.10) & 0.921 (0.14) & 0.416 (0.11) & 1.395 (0.28) & 0.774 (0.40) \\
& ISIW KIE COV & 1.073 (0.18) & 0.420 (0.10) & 1.096 (0.19) & 0.420 (0.11) & 1.737 (0.39) & 0.842 (0.42) \\
\bottomrule
\end{tabular}}
\end{table}

\begin{table}[!ht]
\centering
\caption{Bias (SE) and RMSE (SE) for estimation of $\mu$ across the $B=500$
simulations for the point process misspecification simulation with $n=100$ points. For bias, bold
indicates the smallest value in absolute magnitude within each point process and simulation scenario; for RMSE, bold indicates the smallest value.}
\label{tab:synthetic_mu_bias_rmse}
\resizebox{\textwidth}{!}{
\begin{tabular}{lll cccc cc}
\toprule
\textbf{Metric} & \textbf{Point process} & \textbf{Method} &
\multicolumn{2}{c}{$\boldsymbol{\beta} = -1$} &
\multicolumn{2}{c}{$\boldsymbol{\beta} = 1$} &
\multicolumn{2}{c}{$\boldsymbol{\beta} = 2$} \\
\cmidrule(lr){4-5} \cmidrule(lr){6-7} \cmidrule(lr){8-9}
& & & $\phi = 0.02$ & $\phi = 0.15$ &
         $\phi = 0.02$ & $\phi = 0.15$ &
         $\phi = 0.02$ & $\phi = 0.15$ \\
\midrule
%%%%%%%%%%%%%%%%%%%%%%%%%%%%%%%%%%%%%%%%%%%%%%%%%%%%%%%%%%%%%%%%%%%%%%%%%%%%
%%%%%%%%%%%%%%%%%%%%%%%%%%%%%%% RELATIVE BIAS %%%%%%%%%%%%%%%%%%%%%%%%%%%%%%
%%%%%%%%%%%%%%%%%%%%%%%%%%%%%%%%%%%%%%%%%%%%%%%%%%%%%%%%%%%%%%%%%%%%%%%%%%%%
\multirow{10}{*}{Bias}
& \multirow{5}{*}{LGCP}
& MLE           & -1.278 (0.16) & -0.451 (0.51)
                & 1.267 (0.16) & 0.445 (0.49) & 2.229 (0.19) & 1.052 (0.57) \\
& & INLA-SLP    & -0.395 (0.28) & \textbf{-0.017 (0.50)}
                & 0.368 (0.27)  & \textbf{0.016 (0.50)} & 0.604 (0.35) & 0.115 (0.56) \\
& & ISIW Known  & \textbf{0.084 (0.41)} & 0.377 (4.43)
                & \textbf{-0.088 (0.39)} & -0.600 (0.65)
                & \textbf{0.532 (0.55)} & \textbf{0.067 (0.76)} \\
& & ISIW KIE    & -0.964 (0.45) & 0.230 (1.00)
                & 0.891 (0.40)  & -0.261 (1.23) & 1.713 (2.33) & 0.534 (0.90) \\
& & ISIW KIE COV& -1.204 (0.17) & 0.100 (0.61)
                & 1.193 (0.10)  & -0.105 (0.62) & 2.137 (0.20) & 0.636 (0.66) \\
\cmidrule(lr){2-9}
& \multirow{5}{*}{Thomas}
& MLE           & -1.174 (0.18) & -0.440 (0.50)
                & 1.169 (0.16) & 0.437 (0.50) & 1.853 (0.29) & 0.973 (0.61) \\
& & INLA-SLP    & 0.378 (0.85) & 0.349 (0.41)
                & -0.574 (0.47) & -0.339 (0.38)
                & -2.841 (2.01) & -0.801 (0.90) \\
& & ISIW Known  & \textbf{0.129 (0.39)} & 0.599 (0.64)
                & \textbf{-0.188 (0.40)} & -0.517 (1.24)
                & \textbf{0.457 (0.63)} & \textbf{0.016 (0.80)} \\
& & ISIW KIE    & -0.903 (0.28) & 0.148 (0.83)
                & 0.862 (0.46)  & -0.174 (1.06)
                & 1.698 (0.93) & 0.583 (0.78) \\
& & ISIW KIE COV& -1.094 (0.19) & \textbf{0.087 (0.61)}
                & 1.090 (0.19) & \textbf{-0.109 (0.64)}
                & 1.719 (0.05) & 0.050 (9.48) \\
\midrule
%%%%%%%%%%%%%%%%%%%%%%%%%%%%%%%%%%%%%%%%%%%%%%%%%%%%%%%%%%%%%%%%%%%%%%%%%%%%
%%%%%%%%%%%%%%%%%%%%%%%%%%%%%%% RELATIVE RMSE %%%%%%%%%%%%%%%%%%%%%%%%%%%%%%
%%%%%%%%%%%%%%%%%%%%%%%%%%%%%%%%%%%%%%%%%%%%%%%%%%%%%%%%%%%%%%%%%%%%%%%%%%%%
\multirow{10}{*}{RMSE}
& \multirow{5}{*}{LGCP}
& MLE           & 1.278 (0.16) & 0.555 (0.40)
                & 1.267 (0.16) & 0.546 (0.38) & 2.229 (0.19) & 1.061 (0.55) \\
& & INLA-SLP    & 0.412 (0.26) & \textbf{0.410 (0.29)}
                & 0.388 (0.24) & \textbf{0.410 (0.29)}
                & \textbf{0.623 (0.32)} & \textbf{0.460 (0.34)} \\
& & ISIW Known  & \textbf{0.292 (0.30)} & 0.900 (4.35)
                & \textbf{0.288 (0.28)} & 0.718 (0.52)
                & 0.656 (0.59) & 0.595 (0.47) \\
& & ISIW KIE    & 0.975 (0.43) & 0.582 (0.85)
                & 0.915 (0.34) & 0.612 (1.10)
                & 1.915 (2.16) & 0.743 (0.74) \\
& & ISIW KIE COV& 1.204 (0.17) & 0.483 (0.38)
                & 1.193 (0.16) & 0.489 (0.36)
                & 2.137 (0.20) & 0.752 (0.52) \\
\cmidrule(lr){2-9}
& \multirow{5}{*}{Thomas}
& MLE           & 1.174 (0.18) & 0.542 (0.39)
                & 1.169 (0.16) & 0.536 (0.39)
                & 1.853 (0.29) & 0.983 (0.59) \\
& & INLA-SLP    & 0.717 (0.58) & \textbf{0.416 (0.34)}
                & 0.615 (0.41) & \textbf{0.405 (0.31)}
                & 2.855 (1.99) & 0.914 (0.78) \\
& & ISIW Known  & \textbf{0.296 (0.28)} & 0.717 (0.50)
                & \textbf{0.335 (0.31)} & 0.733 (1.13)
                & \textbf{0.630 (0.46)} & \textbf{0.588 (0.54)} \\
& & ISIW KIE    & 0.912 (0.25) & 0.564 (0.92)
                & 0.912 (0.52) & 0.529 (0.49)
                & 1.698 (0.93) & 0.775 (0.78) \\
& & ISIW KIE COV& 1.094 (0.19) & 0.495 (0.36)
                & 1.090 (0.19) & 0.493 (0.37)
                & 1.739 (1.05) & 1.205 (9.40) \\
\bottomrule
\end{tabular}}
\end{table}

\begin{table}[!ht]
\centering
\caption{Mean CRPS (SE) for the multiple Gaussian process simulation experiment.}
\label{tab:synthetic_multipleGP_crps}
\begin{tabular}{lccccc}
\toprule
Point pattern & MLE & INLA-SLP & ISIW Known & ISIW KIE & ISIW KIE COV \\
\midrule
LGCP   & 0.525 (0.06) & 0.508 (0.05) & 0.507 (0.05) & 0.528 (0.10) & 0.531 (0.06) \\
Thomas & 0.540 (0.07) & 0.637 (0.18) & 0.514 (0.05) & 0.530 (0.06) & 0.555 (0.08) \\
\bottomrule
\end{tabular}
\end{table}

%%%%%%%%%%%% Semi-Synthetic Data Results %%%%%%%%%%%%%%%%%%%%%%%%%
\begin{table}[!ht]
\centering
\caption{Mean CRPS (SE) for the real data simulation. }
\label{tab:semi_synthetic_crps}
\resizebox{\textwidth}{!}{
\begin{tabular}{lccccc}
\toprule
Real dataset & MLE & INLA-SLP & ISIW Known & ISIW KIE & ISIW KIE COV \\
\midrule
Galicia moss        & 0.212 (0.10) & 0.131 (0.09) & 0.158 (0.08) & 0.171 (0.08) & 0.193 (0.10) \\
California AQS & 0.437 (0.11) & 0.261 (0.07) & 0.321 (0.13) & 0.399 (0.15) & 0.415 (0.12) \\
\bottomrule
\end{tabular}}
\end{table}

\clearpage

% References
\appendix
\section*{Supplementary Material}
\pagestyle{plain}
\renewcommand{\thetable}{S\arabic{table}}
\renewcommand{\thefigure}{S\arabic{figure}}
\setcounter{table}{0}  % Reset table counter to start at S1
\setcounter{figure}{0}  % Reset figure counter to start at S1
\setcounter{page}{1}
\pagenumbering{arabic}  % or roman, alph, etc.

\section{Additional details on methods}
\subsection{Maximum likelihood estimation}

Justification for geostatistical inference via the MLE derives from spatial large sample theory. Two main frameworks have dominated asymptotics for geostatistical estimators: increasing domain asymptotics, and fixed domain (or infill) asymptotics. {\em Increasing domain asymptotics} assume that as $n\to\infty$, the study region $\mathcal{D}$ expands, ensuring a minimum separation distance between observation locations and a fixed density of observations per unit area. {\em Fixed domain asymptotics}, on the other hand, keep $\mathcal{D}$ fixed while increasing $n$, leading to a growing observation density and vanishing minimum distance between observations (i.e., observations occur closer together as the sample size increases within the fixed study area).

Under increasing domain asymptotics, the MLE for $\boldsymbol{\psi}$ is consistent and asymptotically normal (AN) \citep{mardia_maximum_1984, bachoc_asymptotic_2014, bachoc_asymptotic_2020}. Results are more challenging under fixed domain asymptotics due to the inclusion of increasingly close observations of a spatially correlated process. Specifically, for a Mat\'ern Gaussian process with known $\nu$, only a subcomponent of the covariance parameters known as the \textit{microergodic} parameter is consistently estimable and asymptotically normal \citep{zhang_inconsistent_2004, kaufman_role_2013}. We refer to this parameter as $\kappa$, given by $\sigma^2 / \phi^{2\nu}$ for Mat\'ern covariance functions. The nugget $\tau^2$ is also estimable \citep{tang_identifiability_2021} while the variance $\sigma^2$ and range $\phi$ are not \citep{zhang_inconsistent_2004}. Although the smoothness parameter $\nu$ is theoretically estimable, it is numerically challenging to estimate and is conventionally treated as fixed \citep{loh_fixed-domain_2021}. Fixed effect coefficients in the mean are typically non-estimable \citep{wang_prediction_2020}, except under specific smoothness conditions on the covariates \citep{yu_parametric_2022, bolin_spatial_2024, gilbert_consistency_2025}. 

\subsection{Composite likelihood}
Let $\textbf{Y}=(Y_1,Y_2,...,Y_n)^\top\in \mathbb{R}^n$ be a $n\times 1$ random vector with probability density $f(\textbf{y};\boldsymbol{\psi})$ for unknown $r$-dimensional parameter vector $\boldsymbol{\psi}\in \Psi\subseteq \mathbb{R}^r$. Define $\{\mathcal{A}_1,...,\mathcal{A}_K\}$ to be a set of $K$ marginal or conditional events with associated likelihood $\mathcal{L}_k(\theta;\textbf{y})\propto f(\textbf{y}\in\mathcal{A}_k;\boldsymbol{\psi})$. The log composite likelihood (CL) is the weighted sum of the event-specific log likelihoods
\begin{equation*}\label{WCL}
    \log \mathcal{L}_C(\boldsymbol{\psi};\textbf{y}):= \sum_{k=1}^K w_k \log\mathcal{L}_k(\boldsymbol{\psi};\textbf{y}),
\end{equation*}
where $\textbf{w}:=(w_1,...,w_K)$ is a vector of weights, not necessarily non-negative and the density $f$ follows that defined in \eqref{eqn:aim2_MLE_likelihood}. The maximum CL estimate is defined as $\hat{\boldsymbol\psi}_{C}:=\text{argmax}_{\boldsymbol{\psi}} \log \mathcal{L}_C(\boldsymbol{\psi};\textbf{y})$. 
One common choice of CL is the univariate marginal where the log likelihood is the sum of the log marginal density of each observation. 
\begin{equation*}
    \log \mathcal{L}_{UM}(\boldsymbol{\psi};\textbf{y})=\sum_{i=1}^n w_i \log f(y_i;\boldsymbol{\psi}).
\end{equation*}

Typically, the use of this univariate marginal in geostatistics is limited because it ignores dependencies between observations and thus cannot estimate covariance parameters. For this reason, pairwise CLs have been much more popular \citep{varin_overview_2011}. \cite{bevilacqua_comparing_2015} compared the efficiency of three different pairwise CL estimators: pairwise marginal (PMLE), pairwise conditional (PCMLE), and pairwise difference (PDMLE) CLs. These authors concluded that the PMLE outperformed all other pairwise CLs and recommended its use over the PCMLE and PDMLE. The likelihoods for each pairwise CL are
\begin{equation} \label{eqn:aim2_PCL}
\begin{split}
    \log \mathcal{L}_{PM}(\boldsymbol{\psi};\textbf{y}) = \sum_{i<j} w_{ij} \log f(y_i, y_j;\boldsymbol{\psi}), \\
    \log \mathcal{L}_{PC}(\boldsymbol{\psi};\textbf{y}) = \sum_{i\neq j} w_{ij} \log f(y_i|y_j;\boldsymbol{\psi}), \\
    \log \mathcal{L}_{PD}(\boldsymbol{\psi};\textbf{y}) = \sum_{i<j} w_{ij} \log f(y_i-y_j;\boldsymbol{\psi}). 
\end{split}
\end{equation}

% Similar to the MLE, asymptotic results for maximum CL estimators for independent and identically distributed $\textbf{Y}$ are available under standard regularity conditions \citep{varin_overview_2011}. These typically require the likelihood to meet the Bartlett identity conditions {\bf (add citation to these)}. If these conditions are met, \eqref{WCL} is an unbiased estimating equation and under regularity conditions, $\hat{\boldsymbol{\psi}}_C$ is consistent and asymptotically normal. The maximum CL estimator is asymptotically normal with the following covariance matrix.
% \begin{equation}
%     \begin{split}
%         \textbf{V}(\boldsymbol{\psi}_0)&=n^{-1} \textbf{A}^{-1}(\boldsymbol{\psi}_0)\textbf{B}(\boldsymbol{\psi}_0)\textbf{A}^{-1}(\boldsymbol{\psi}_0), \\
%         \textbf{A}(\boldsymbol{\psi}) &= \mathbb{E}\left(- \frac{\partial^2}{\partial\boldsymbol{\psi}\partial \boldsymbol{\psi}^\top}\log \mathcal{L}_C(\boldsymbol{\psi};\textbf{y}) \right),\\
%         \textbf{B}(\boldsymbol{\psi})&=\text{Var}\left( \frac{\partial}{\partial\boldsymbol{\psi}} \log\mathcal{L}_C(\boldsymbol{\psi};\textbf{y})\right).
%     \end{split}
% \end{equation}

% Once again, these standard results do not apply under \eqref{MLE} due to the unique dependence enforced by the covariance function $C(\cdot, \cdot)$ and require a different set of theoretical tools. 

These authors also proved, under increasing domain asymptotics, the consistency and asymptotic normality of $\boldsymbol{\hat{\psi}}_C$ estimated from \eqref{eqn:aim2_PCL}. \cite{bachoc_composite_2019} proved the same properties for the PMLE and PCMLE in the one-dimensional setting under fixed domain asymptotics. However, the true utility of CL lies in its computational speed and robustness to misspecification. While the MLE requires a $O(n^3)$ matrix inversion and specification of the joint density, pairwise CL consists only of $O(n^2)$ terms and only requires correct specification of second order densities. The computational efficiency of pairwise CL can be further enhanced by using the weights $\textbf{w}$ to only include pairwise observations within a certain distance $d$ apart \citep{bevilacqua_comparing_2015}.

In this study, we use the PMLE as a basis for ISIW due to its superior performance over other pairwise likelihoods. In particular, we do not consider the univariate marginal or PDMLE because they do not directly estimate the entirety of $\boldsymbol\psi$. 

\subsection{Vecchia approximation}

Theoretical justification of the Vecchia approximation in spatial statistics is relatively light compared to that of the MLE and CL. Typical approaches to proving fixed domain asymptotics fail due to the loss of stationarity in the parent process introduced by the Vecchia approximation \citep{zhang_fixed-domain_2024}. However, it is impossible to deny its excellent empirical performance in applications \citep{heaton_case_2019}. A key advantage the Vecchia approximation possesses over typical CL is that \eqref{eqn:aim2_vecchia} and \eqref{eqn:aim2_vecchia_CL} correspond to a valid joint probability distribution for $\textbf{Y}$. As a result, the Kullback-Leibler (KL) divergence of the Vecchia approximation with respect to the true distribution can be computed and has been shown to be a nonincreasing function of $m$ when the conditioning sets $q(i)$ are chosen as the nearest neighbors \citep{katzfuss_vecchia_2020, huser_vecchia_2023}.

\subsection{Kriging}

Stein established the asymptotic efficiency of kriging under various robustness conditions \citep{stein_asymptotically_1988, stein_uniform_1990, stein_bounds_1990, stein_simple_1993}, while \cite{wang_prediction_2020} showed that kriging's prediction error vanishes under a uniform metric. Additionally, \cite{putter_effect_2001} demonstrated that the difference between predictions using kriging with estimated and the true parameters is asymptotically negligible if the joint Gaussian distributions of the spatial process under the true and estimated covariance functions are contiguous almost surely. Given these results, one could argue that Gaussian process models are more appropriate for spatial interpolation rather than inference.

\section{Additional simulation results}

\subsection{Choice of kernel estimator}\label{sec:KIE_choice}

To select a representative version of the ISIW estimator and improve the interpretability of our simulation, we first conducted a preliminary analysis before the main simulation study. The ISIW estimator requires selecting a composite likelihood and a kernel intensity estimator (KIE) for the weights. We investigated two likelihoods: the pairwise marginal composite likelihood (PMLE) and Vecchia approximation. We denote their ISIW versions by ISIW-PM and ISIW-V respectively.

We generated $B=500$ Monte Carlo simulations for 12 distinct scenarios based on the SLP model (with fixed parameters $\mu=4$, $\sigma^2=1.5$, $\tau^2=0.1$, and $\beta=1$). These scenarios varied based on:

\begin{itemize}
    \item Sample size of the observation process ($N=100, 800$).
    \item Strength of spatial correlation ($\phi=0.02, 0.15$).
    \item Parametric form of the underlying point process ($\mathbf{X}$ follows a LGCP, sigmoidal Cox process, or Thomas processs driven by $S$)
\end{itemize}

We then ranked each CL-KIE combination based on its RMSPE when predicting the spatial field on a $32 \times 32$ grid. All considered KIEs and their bandwidth selection strategies are listed in Table \ref{tab:aim2_bw_selection}. Based on the RMSPE rankings compiled across these 12 scenarios (Table \ref{tab:aim2_rank}), we determined that the ISIW combination of Vecchia approximation and the \texttt{CvL.adaptive} bandwidth selection yielded the best results. 

\begin{table}[h]
    \centering
    \renewcommand{\arraystretch}{1.2} % Adjust row height for better readability
    \caption{Point process intensity function estimators considered in the simulation study.}
    \resizebox{\textwidth}{!}{  % Ensures table fits within page width
        \begin{tabular}{l l l}  
            \toprule
            \textbf{Method} & \textbf{Reference} & \textbf{Bandwidth Selection} \\
            \midrule
            \texttt{diggle} & \cite{diggle_kernel_1985} & Least-squares cross-validation \\
            \texttt{scott} & \cite{scott_kernel_1992} & Rule-of-thumb based on normal reference density \\
            \texttt{ppl} & \cite{loader_density_1999} & Likelihood cross-validation (leave-one-out) \\
            \texttt{CvL} & \cite{cronie_non-model-based_2018} & Maximum likelihood cross-validation \\
            \texttt{CvL.adaptive} & \cite{van_lieshout_infill_2021} & Adaptive bandwidth based on local point density \\
            \texttt{R-INLA} & \cite{simpson_going_2016} & LGCP model (non-kernel-based) \\
            \bottomrule
        \end{tabular}
    }
    \label{tab:aim2_bw_selection}
\end{table}

\renewcommand{\arraystretch}{1.1}
\begin{table}[h]
    \centering
    \caption{Predictive performance of all sixteen methods based on median rank, mean rank, and percentage of total simulations when RMSPE for the method was lower than that of MLE and SLP. Rank was determined by RMSPE.}
    \label{tab:method_ranking}
    
    \resizebox{\textwidth}{!}{ % Ensures table width fits within margins
    \begin{tabular}{lcccc}
        \toprule
        Method & Median Rank & Mean Rank & \%  RMSPE lower & \% lower RMSPE \\
         & of RMSPE & of RMSPE & than MLE & than SLP \\
        \midrule
        INLA-SLP & 2 & 5.29 & 73.3 & NA\\
        ISIW-V Known & 2 & 5.75 & 70.0 & 44.0 \\
        ISIW-V CvL.adaptive & 5 & 6.53 & 72.7 & 31.8 \\
        ISIW-V diggle & 5 & 5.96 & 78.5 & 29.3 \\
        ISIW-V CvL & 7 & 7.38 & 85.0 & 28.9 \\
        ISIW-V ppl & 7 & 7.37 & 74.4 & 27.8 \\
        ISIW-V INLA & 8 & 8.23 & 74.0 & 27.9 \\
        ISIW-PM CvL.adaptive & 8 & 8.01 & 65.1 & 28.3 \\
        ISIW-PM diggle & 8 & 7.98 & 61.3 & 28.1 \\
        ISIW-V scott & 9 & 8.29 & 85.6 & 28.4 \\
        ISIW-PM Known & 9 & 8.60 & 51.8 & 23.9 \\
        ISIW-PM ppl & 10 & 9.84 & 50.4 & 27.9 \\
        MLE & 11 & 10.2 & NA & 26.7 \\
        ISIW-PM CvL & 12 & 11.3 & 30.9 & 26.5 \\
        ISIW-PM INLA & 14 & 12.3 & 27.0 & 24.9 \\
        ISIW-PM scott & 14 & 12.7 & 22.3 & 24.6 \\
        \bottomrule
    \end{tabular}
    }
    \label{tab:aim2_rank}
\end{table}

\begin{table}[!ht]
    \centering
    \caption{Relative bias and RMSE in parameter estimation for MLE, SLP, and ISIW methods across all simulation scenarios for preliminary analysis. Bolded values indicate the method with the smallest bias or RMSE for a given parameter.}
    \label{tab:aim2_comparison_param}
    \renewcommand{\arraystretch}{1.2}
    \setlength{\tabcolsep}{5pt}
     \resizebox{\textwidth}{!}{
    \begin{tabular}{llrrrrrrrrrr}
        \toprule
        \multirow{2}{*}{Method} & \multirow{2}{*}{Variant} & \multicolumn{2}{c}{$\mu$} & \multicolumn{2}{c}{$\sigma^2$} & \multicolumn{2}{c}{$\phi$} & \multicolumn{2}{c}{$\tau^2$} & \multicolumn{2}{c}{$\kappa$} \\
        \cmidrule(lr){3-4} \cmidrule(lr){5-6} \cmidrule(lr){7-8} \cmidrule(lr){9-10} \cmidrule(lr){11-12}
        & & Bias & RMSE & Bias & RMSE & Bias & RMSE & Bias & RMSE & Bias & RMSE \\
        \midrule
        MLE & - & 0.106 & 0.173 & 0.046 & 0.589 & \textbf{0.013} & 0.429 & -0.057 & \textbf{0.897} & \textbf{0.113} & \textbf{0.538} \\
        INLA-SLP & - & \textbf{0.016} & \textbf{0.116} & 0.241 & 1.32 & 0.110 & 0.434 & $2.13 \times 10^2$ & $1.19 \times 10^4$ & 0.546 & 17.1 \\[0.8em]

        \multirow{4}{*}{ISIW-V} & Known & -0.259 & 0.409 & 0.035 & 20.8 & 90.9 & $4.52 \times 10^3$ & 0.053 & 1.72 & 0.712 & 1.58 \\
        & CvL & 0.022 & 0.240 & -0.057 & 1.10 & 65.1 & $2.71 \times 10^3$ & \textbf{-0.007} & 1.14 & 0.140 & 0.624 \\
        & CvL.adaptive & -0.018 & 0.268 & 0.362 & 51.9 & $1.25 \times 10^3$ & $3.08 \times 10^4$ & 0.441 & 2.44 & 2.44 & 1.14 \\
        & INLA & 0.059 & 0.170 & -0.068 & 0.807 & -0.023 & 0.673 & -0.013 & 1.15 & 0.166 & 0.804 \\[0.8em]

        \multirow{4}{*}{ISIW-PM} & Known & -0.161 & 0.309 & -0.293 & \textbf{0.408} & -0.352 & 0.489 & 0.713 & 1.09 & 1.900 & 7.65 \\
        & CvL & 0.122 & 0.203 & 0.061 & 0.596 & -0.168 & \textbf{0.397} & 0.871 & 1.50 & 0.885 & 2.05 \\
        & CvL.adaptive & 0.042 & 0.150 & -0.162 & 0.511 & -0.184 & 0.630 & 0.782 & 1.35 & 1.550 & 11.0 \\
        & INLA & 0.135 & 0.211 & \textbf{-0.018} & 0.570 & -0.201 & 0.414 & 0.859 & 1.56 & 0.912 & 1.69 \\
        \bottomrule
    \end{tabular}
    }
\end{table}

\begin{table}[!h]
    \centering
    \caption{Mean (SD) and Median (IQR) runtime in seconds over all simulations for methods under different sample sizes.}
    \label{tab:aim2_runtime_results}
    \resizebox{\textwidth}{!}{%
    \begin{tabular}{ll c c c c}
        \toprule
        & & \multicolumn{2}{c}{N = 100} & \multicolumn{2}{c}{N = 800} \\
        \cmidrule(lr){3-4} \cmidrule(lr){5-6}
        Method & Variant & Mean (SD) & Median (IQR) & Mean (SD) & Median (IQR) \\
        \midrule
        MLE &  & 0.80 (0.26) & 0.76 (0.34) & 43.8 (13.8) & 41.3 (17.2) \\[0.5em]
        INLA-SLP &  & 124.0 (71.8) & 100.0 (87.3) & 99.8 (39.3) & 95.2 (37.3) \\[0.8em]
        
        \multirow{6}{*}{ILIW} & Known & 4.80 (1.81) & 4.47 (2.20) & 45.8 (16.5) & 43.3 (20.9) \\
        & CvL & 4.14 (1.48) & 3.85 (1.89) & 34.4 (10.7) & 32.7 (13.6) \\
        & CvL.adaptive & 4.97 (1.75) & 4.63 (2.11) & 41.6 (13.7) & 39.4 (17.5) \\
        & INLA & 15.9 (4.51) & 15.0 (5.39) & 118.0 (41.4) & 107.0 (51.0) \\
        & diggle & 4.83 (1.93) & 4.44 (2.24) & 42.5 (17.3) & 39.1 (22.1) \\
        & ppl & 4.88 (1.89) & 4.49 (2.16) & 39.3 (16.2) & 36.0 (16.8) \\
        & scott & 4.08 (1.44) & 3.79 (1.83) & 34.3 (11.0) & 32.6 (13.8) \\[0.8em]
        
        \multirow{6}{*}{PMLE} & Known & 0.26 (0.06) & 0.25 (0.06) & 16.5 (2.4) & 16.2 (3.3) \\
        & CvL & 0.25 (0.05) & 0.24 (0.05) & 14.9 (2.1) & 14.8 (2.8) \\
        & CvL.adaptive & 0.59 (0.14) & 0.55 (0.15) & 16.6 (2.1) & 16.4 (2.9) \\
        & INLA & 9.61 (3.90) & 8.40 (3.41) & 97.4 (36.4) & 86.3 (41.8) \\
        & diggle & 0.28 (0.06) & 0.27 (0.06) & 16.4 (2.3) & 16.2 (3.2) \\
        & ppl & 0.65 (0.13) & 0.64 (0.18) & 15.7 (2.1) & 15.6 (2.8) \\
        & scott & 0.25 (0.05) & 0.25 (0.05) & 15.0 (2.3) & 14.8 (3.3) \\[0.8em]
        
        \bottomrule
    \end{tabular}%
    }
\end{table}

\clearpage
\subsection{Sensitivity analysis for winsorization threshold}\label{sec:Winsorization_choice}

Once a representative KIE estimator was chosen, we conducted a sensitivity analysis of the choice of threshold for the winsorization to mitigate any numerical issues caused by extreme weights in the likelihood adjustment. Percentiles from 90\% to 99\% were considered. We ran $B=100$ Monte Carlo simulations under the random field specifications outlined in the point process misspecification simulation portion of the main text, restricted to the SLP model with parameters $(\mu, \sigma^2, \phi, \nu, \tau^2)=(4, 1.5, 0.15, 1, 0.1)$  and $\beta=1$ or $\beta=2$. Based on using both RMSPE and CRPS as evaluation metrics, we chose 93\% to use for the remainder of the study.

\begin{table}[h]
\centering
\caption{RMSPE (SE) of ISIW KIE Estimator under different winsorization thresholds. The NA cell indicates there were simulation runs that did not converge due to numerical instability. For the $\beta=2$ scenario three out of 100 failed while for $\beta=1$ scenario two out of 100 failed.}
\label{tab:winsorization_sensitivity_rmspe}
\begin{tabular}{lcc}
\toprule
\textbf{} & \multicolumn{2}{c}{\textbf{RMSPE (SE)}} \\
\cmidrule(lr){2-3}
\textbf{(\% Upper Quantile)} & ${\beta}=2$ & ${\beta}=1$ \\
\midrule
90\% & 1.242 (0.364) & 0.741 (0.176) \\ 
91\% & 1.233 (0.365) & 0.755 (0.281) \\ 
92\% & 1.221 (0.364) & 0.750 (0.280) \\ 
93\% & 1.214 (0.369) & 0.749 (0.281) \\ 
94\% & 1.212 (0.375) & 0.767 (0.314) \\ 
95\% & 1.201 (0.380) & 0.762 (0.310) \\ 
96\% & 1.203 (0.397) & 0.789 (0.375) \\ 
97\% & 1.221 (0.473) & 0.783 (0.340) \\ 
98\% & 1.247 (0.434) & 0.821 (0.382) \\ 
99\% & \text{NA} & \text{NA} \\ 
\bottomrule
\end{tabular}
\end{table}

\begin{table}[h]
\centering
\caption{CRPS (SE) of ISIW KIE Estimator under different winsorization thresholds. Same format as Table \ref{tab:winsorization_sensitivity_rmspe}.}
\label{tab:winsorization_sensitivity_crps}
\begin{tabular}{lcc}
\toprule
\textbf{} & \multicolumn{2}{c}{\textbf{CRPS (SE)}} \\
\cmidrule(lr){2-3}
\textbf{(\% Upper Quantile)} & ${\beta}=2$ & ${\beta}=1$ \\
\midrule
90\% & 0.755 (0.278) & 0.394 (0.095) \\ 
91\% & 0.750 (0.278) & 0.413 (0.247) \\ 
92\% & 0.741 (0.277) & 0.411 (0.248) \\ 
93\% & 0.741 (0.286) & 0.415 (0.249) \\ 
94\% & 0.744 (0.304) & 0.435 (0.285) \\ 
95\% & 0.744 (0.314) & 0.435 (0.284) \\ 
96\% & 0.763 (0.345) & 0.466 (0.354) \\ 
97\% & 0.802 (0.445) & 0.467 (0.318) \\ 
98\% & 0.855 (0.421) & 0.510 (0.360) \\ 
99\% & \text{NA} & \text{NA} \\ 
\bottomrule
\end{tabular}
\end{table}

\clearpage
\subsection{Simulation experiment}

%%%%%%%%%%%%%%%Sigma2%%%%%%%%%%%%%%%%%%%%%%%%%%
\begin{table}[h]
\centering
\caption{Bias (SE) and RMSE (SE) for estimation of $\sigma^2$ across the $B=500$
simulations for the point process misspecification simulation experiment with $N=100$ points. For bias, bold
indicates the smallest value in absolute magnitude within each point process and simulation scenario; for RMSE, bold indicates the smallest value. The NA* indicates some simulations diverged.}
\label{tab:sigma2_bias_rmse}
\resizebox{\textwidth}{!}{
\begin{tabular}{lll cccc cc}
\toprule
\textbf{Metric} & \textbf{Point process} & \textbf{Method} &
\multicolumn{2}{c}{$\boldsymbol{\beta} = -1$} &
\multicolumn{2}{c}{$\boldsymbol{\beta} = 1$} &
\multicolumn{2}{c}{$\boldsymbol{\beta} = 2$} \\
\cmidrule(lr){4-5} \cmidrule(lr){6-7} \cmidrule(lr){8-9}
& & & $\phi = 0.02$ & $\phi = 0.15$ &
         $\phi = 0.02$ & $\phi = 0.15$ &
         $\phi = 0.02$ & $\phi = 0.15$ \\
\midrule
%%%%%%%%%%%%%%%%%%%%%%%%%%%%%%%%%%%%%%%%%%%%%%%%%%%%%%%%%%%%%%%%%%%%%%%%%%%%
%%%%%%%%%%%%%%%%%%%%%%%%%%%%%%% BIAS %%%%%%%%%%%%%%%%%%%%%%%%%%%%%%%%%%%%%%%
%%%%%%%%%%%%%%%%%%%%%%%%%%%%%%%%%%%%%%%%%%%%%%%%%%%%%%%%%%%%%%%%%%%%%%%%%%%%
\multirow{10}{*}{Bias}
& \multirow{5}{*}{LGCP}
& MLE           & -0.156 (0.28) & -0.461 (0.41)
                & \textbf{-0.163 (0.26)} & -0.453 (0.43)
                & -0.581 (0.18) & -0.846 (0.29) \\
& & INLA-SLP    & \textbf{-0.022 (0.38)} & \textbf{-0.423 (0.38)}
                & NA* & \textbf{-0.423 (0.41)}
                & \textbf{-0.167 (0.39)} & \textbf{-0.633 (0.34)} \\
& & ISIW Known  & -0.511 (0.41) & -0.511 (6.95)
                & -0.523 (0.40) & -0.808 (0.69)
                & -1.097 (0.26) & -1.196 (0.17) \\
& & ISIW KIE    & -0.397 (0.46) & -0.822 (0.49)
                & -0.451 (0.47) & 1.338 (47.86)
                & -0.649 (2.04) & -1.061 (0.30) \\
& & ISIW KIE COV& -0.916 (0.24) & -0.850 (0.27)
                & -0.968 (0.28) & -0.826 (0.34)
                & -1.260 (0.18) & -1.235 (0.16) \\
\cmidrule(lr){2-9}
& \multirow{5}{*}{Thomas}
& MLE           & \textbf{-0.181 (0.24)} & -0.434 (0.47)
                & \textbf{-0.155 (0.27)} & -0.439 (0.46)
                & \textbf{-0.584 (0.26)} & -0.746 (0.54) \\
& & INLA-SLP    & 13.357 (277.27) & \textbf{-0.324 (0.58)}
                & 0.506 (0.73) & \textbf{-0.342 (0.49)}
                & 6.345 (46.20) & \textbf{-0.380 (0.89)} \\
& & ISIW Known  & -0.524 (0.37) & -0.854 (0.26)
                & -0.534 (0.40) & -0.785 (1.92)
                & -1.124 (0.42) & -1.232 (0.17) \\
& & ISIW KIE    & -0.290 (0.39) & 0.847 (27.45)
                & -0.270 (0.45) & -0.648 (2.28)
                & -0.587 (0.34) & -0.895 (0.52) \\
& & ISIW KIE COV& -1.008 (0.25) & -0.851 (0.36)
                & -1.026 (0.27) & -0.882 (0.33)
                & -1.276 (0.22) & -0.880 (7.03) \\
\midrule
%%%%%%%%%%%%%%%%%%%%%%%%%%%%%%%%%%%%%%%%%%%%%%%%%%%%%%%%%%%%%%%%%%%%%%%%%%%%
%%%%%%%%%%%%%%%%%%%%%%%%%%%%%%% RMSE %%%%%%%%%%%%%%%%%%%%%%%%%%%%%%%%%%%%%%%
%%%%%%%%%%%%%%%%%%%%%%%%%%%%%%%%%%%%%%%%%%%%%%%%%%%%%%%%%%%%%%%%%%%%%%%%%%%%
\multirow{10}{*}{RMSE}
& \multirow{5}{*}{LGCP}
& MLE           & \textbf{0.249 (0.20)} & 0.548 (0.28)
                & \textbf{0.241 (0.19)} & 0.546 (0.30)
                & 0.582 (0.18) & 0.857 (0.26) \\
& & INLA-SLP    & 0.268 (0.26) & \textbf{0.502 (0.27)}
                & NA* & \textbf{0.515 (0.29)}
                & \textbf{0.339 (0.26)} & \textbf{0.662 (0.28)} \\
& & ISIW Known  & 0.571 (0.32) & 1.172 (6.87)
                & 0.573 (0.33) & 0.870 (0.61)
                & 1.100 (0.25) & 1.196 (0.17) \\
& & ISIW KIE    & 0.504 (0.34) & 0.879 (0.38)
                & 0.537 (0.37) & 3.043 (47.78)
                & 0.937 (1.92) & 1.074 (0.25) \\
& & ISIW KIE COV& 0.916 (0.24) & 0.862 (0.23)
                & 0.968 (0.28) & 0.859 (0.25)
                & 1.260 (0.18) & 1.235 (0.16) \\
\cmidrule(lr){2-9}
& \multirow{5}{*}{Thomas}
& MLE           & \textbf{0.245 (0.17)} & 0.556 (0.32)
                & \textbf{0.255 (0.18)} & 0.561 (0.29)
                & \textbf{0.594 (0.24)} & 0.846 (0.36) \\
& & INLA-SLP    & 13.632 (277.26) & \textbf{0.524 (0.41)}
                & 0.651 (0.60) & \textbf{0.513 (0.31)}
                & 6.383 (46.20) & \textbf{0.758 (0.60)} \\
& & ISIW Known  & 0.569 (0.29) & 0.862 (0.23)
                & 0.588 (0.32) & 0.967 (1.83)
                & 1.155 (0.32) & 1.232 (0.17) \\
& & ISIW KIE    & 0.397 (0.28) & 2.357 (27.36)
                & 0.421 (0.32) & 0.915 (2.19)
                & 0.624 (0.27) & 0.971 (0.35) \\
& & ISIW KIE COV& 1.008 (0.25) & 0.875 (0.29)
                & 1.026 (0.27) & 0.906 (0.26)
                & 1.282 (0.18) & 1.551 (6.92) \\
\bottomrule
\end{tabular}}
\end{table}

%%%%%%%%%%%%%%%%% Range %%%%%%%%%%%%%%%%%%%%%%%%%%%%%%
\begin{table}[h]
\centering
\caption{Bias (SE) and RMSE (SE) for estimation of the range parameter $\phi$ across the $B=500$
simulations for the point process misspecification simulation experiment with $N=100$ points. For bias, bold
indicates the smallest value in absolute magnitude within each point process and simulation
scenario; for RMSE, bold indicates the smallest value.}
\label{tab:range_bias_rmse}
\resizebox{\textwidth}{!}{
\begin{tabular}{lll cccc cc}
\toprule
\textbf{Metric} & \textbf{Point process} & \textbf{Method} &
\multicolumn{2}{c}{$\boldsymbol{\beta} = -1$} &
\multicolumn{2}{c}{$\boldsymbol{\beta} = 1$} &
\multicolumn{2}{c}{$\boldsymbol{\beta} = 2$} \\
\cmidrule(lr){4-5} \cmidrule(lr){6-7} \cmidrule(lr){8-9}
& & & $\phi = 0.02$ & $\phi = 0.15$ &
         $\phi = 0.02$ & $\phi = 0.15$ &
         $\phi = 0.02$ & $\phi = 0.15$ \\
\midrule
%%%%%%%%%%%%%%%%%%%%%%%%%%%%%%%%%%%%%%%%%%%%%%%%%%%%%%%%%%%%%%%%%%%%%%%%%%%%
%%%%%%%%%%%%%%%%%%%%%%%%%%%%%%% BIAS %%%%%%%%%%%%%%%%%%%%%%%%%%%%%%%%%%%%%%%
%%%%%%%%%%%%%%%%%%%%%%%%%%%%%%%%%%%%%%%%%%%%%%%%%%%%%%%%%%%%%%%%%%%%%%%%%%%%
\multirow{10}{*}{Bias}
& \multirow{5}{*}{LGCP}
& MLE           & \textbf{0.001 (0.01)} & -0.032 (0.04)
                & \textbf{0.000 (0.01)} & -0.032 (0.04)
                & \textbf{-0.004 (0.01)} & -0.061 (0.03) \\
& & INLA-SLP    & NA* & \textbf{-0.026 (0.03)}
                & NA* & \textbf{-0.026 (0.03)}
                & 0.005 (0.01) & \textbf{-0.026 (0.03)} \\
& & ISIW Known  & 0.008 (0.05) & -0.063 (0.09)
                & 5.998 (133.91) & -0.065 (0.03)
                & 201.968 (2406.95) & 0.068 (2.57) \\
& & ISIW KIE    & 207.207 (4627.34) & -0.048 (0.15)
                & 78.978 (968.22) & 131.120 (2709.06)
                & NA* & NA* \\
& & ISIW KIE COV& NA* & -0.060 (0.04)
                & NA* & -0.057 (0.04)
                & NA* & 95.741 (1887.07) \\
\cmidrule(lr){2-9}
& \multirow{5}{*}{Thomas}
& MLE           & \textbf{-0.002 (0.01)} & \textbf{-0.032 (0.04)}
                & \textbf{-0.001 (0.01)} & \textbf{-0.029 (0.04)}
                & -0.008 (0.00) & -0.056 (0.06) \\
& & INLA-SLP    & 0.045 (0.64) & -0.046 (0.02)
                & 0.010 (0.01) & -0.044 (0.03)
                & 0.028 (0.01) & \textbf{-0.050 (0.02)} \\
& & ISIW Known  & 0.002 (0.04) & -0.066 (0.03)
                & 2.261 (41.64) & -0.062 (0.05)
                & 18.463 (182.27) & 9.139 (205.42) \\
& & ISIW KIE    & 2.029 (45.22) & 0.044 (1.47)
                & 16.706 (226.19) & -0.031 (0.20)
                & \textbf{-0.006 (0.02)} & 13.117 (212.18) \\
& & ISIW KIE COV& NA* & -0.059 (0.05)
                & NA* & -0.056 (0.05)
                & NA* & 13.713 (226.37) \\
\midrule
%%%%%%%%%%%%%%%%%%%%%%%%%%%%%%%%%%%%%%%%%%%%%%%%%%%%%%%%%%%%%%%%%%%%%%%%%%%%
%%%%%%%%%%%%%%%%%%%%%%%%%%%%%%% RMSE %%%%%%%%%%%%%%%%%%%%%%%%%%%%%%%%%%%%%%%
%%%%%%%%%%%%%%%%%%%%%%%%%%%%%%%%%%%%%%%%%%%%%%%%%%%%%%%%%%%%%%%%%%%%%%%%%%%%
\multirow{10}{*}{RMSE}
& \multirow{5}{*}{LGCP}
& MLE           & \textbf{0.005 (0.01)} & 0.043 (0.03)
                & \textbf{0.004 (0.00)} & 0.043 (0.03)
                & \textbf{0.006 (0.00)} & 0.064 (0.03) \\
& & INLA-SLP    & NA* & \textbf{0.035 (0.02)}
                & NA* & \textbf{0.035 (0.02)}
                & 0.007 (0.01) & \textbf{0.032 (0.02)} \\
& & ISIW Known  & 0.014 (0.04) & 0.073 (0.08)
                & 6.004 (133.91) & 0.069 (0.03)
                & 201.975 (2406.95) & 0.224 (2.56) \\
& & ISIW KIE    & 207.211 (4627.34) & 0.079 (0.14)
                & 78.982 (968.22) & 131.239 (2709.05)
                & NA* & NA* \\
& & ISIW KIE COV& NA* & 0.065 (0.03)
                & NA* & 0.066 (0.03)
                & NA* & 95.910 (1887.06) \\
\cmidrule(lr){2-9}
& \multirow{5}{*}{Thomas}
& MLE           & \textbf{0.004 (0.00)} & \textbf{0.043 (0.03)}
                & \textbf{0.004 (0.00)} & \textbf{0.042 (0.03)}
                & \textbf{0.009 (0.00)} & 0.070 (0.05) \\
& & INLA-SLP    & 0.045 (0.64) & 0.047 (0.02)
                & 0.011 (0.01) & 0.046 (0.02)
                & 0.028 (0.01) & \textbf{0.050 (0.02)} \\
& & ISIW Known  & 0.010 (0.03) & 0.069 (0.03)
                & 2.268 (41.64) & 0.070 (0.04)
                & 18.478 (182.27) & 9.307 (205.41) \\
& & ISIW KIE    & 2.035 (45.22) & 0.157 (1.46)
                & 16.711 (226.19) & 0.077 (0.19)
                & 0.010 (0.01) & 13.259 (212.17) \\
& & ISIW KIE COV& NA* & 0.067 (0.03)
                & NA* & 0.065 (0.03)
                & NA* & 13.874 (226.36) \\
\bottomrule
\end{tabular}}
\end{table}

%%%%%%%%%%%%%%%%% Nugget %%%%%%%%%%%%%%%%%%%%%%%%%%%
\begin{table}[h]
\centering
\caption{Bias (SE) and RMSE (SE) for estimation of the nugget across the $B=500$
simulations for the point process misspecification simulation experiment with $N=100$ points. For bias, bold
indicates the smallest value in absolute magnitude within each point process and simulation
scenario; for RMSE, bold indicates the smallest value.}
\label{tab:nugget_bias_rmse}
\resizebox{\textwidth}{!}{
\begin{tabular}{lll cccc cc}
\toprule
\textbf{Metric} & \textbf{Point process} & \textbf{Method} &
\multicolumn{2}{c}{$\boldsymbol{\beta} = -1$} &
\multicolumn{2}{c}{$\boldsymbol{\beta} = 1$} &
\multicolumn{2}{c}{$\boldsymbol{\beta} = 2$} \\
\cmidrule(lr){4-5} \cmidrule(lr){6-7} \cmidrule(lr){8-9}
& & & $\phi = 0.02$ & $\phi = 0.15$ &
         $\phi = 0.02$ & $\phi = 0.15$ &
         $\phi = 0.02$ & $\phi = 0.15$ \\
\midrule
%%%%%%%%%%%%%%%%%%%%%%%%%%%%%%%%%%%%%%%%%%%%%%%%%%%%%%%%%%%%%%%%%%%%%%%%%%%%
%%%%%%%%%%%%%%%%%%%%%%%%%%%%%%% BIAS %%%%%%%%%%%%%%%%%%%%%%%%%%%%%%%%%%%%%%%
%%%%%%%%%%%%%%%%%%%%%%%%%%%%%%%%%%%%%%%%%%%%%%%%%%%%%%%%%%%%%%%%%%%%%%%%%%%%
\multirow{10}{*}{Bias}
& \multirow{5}{*}{LGCP}
& MLE           & \textbf{0.002 (0.15)} & \textbf{-0.007 (0.03)}
                & \textbf{-0.004 (0.13)} & -0.008 (0.03)
                & -0.019 (0.07) & -0.008 (0.03) \\
& & INLA-SLP    & -0.048 (0.49) & 7.426 (166.39)
                & NA* & -0.016 (0.04)
                & \textbf{0.009 (0.51)} & \textbf{-0.003 (0.03)} \\
& & ISIW Known  & 0.066 (0.25) & -0.015 (0.05)
                & 0.088 (0.30) & -0.018 (0.05)
                & 0.069 (0.18) & -0.013 (0.05) \\
& & ISIW KIE    & 0.109 (0.27) & -0.018 (0.07)
                & 0.133 (0.31) & \textbf{-0.005 (0.10)}
                & 0.091 (0.22) & -0.008 (0.05) \\
& & ISIW KIE COV& 0.070 (0.20) & -0.010 (0.05)
                & 0.105 (0.24) & -0.013 (0.04)
                & 0.087 (0.17) & -0.012 (0.05) \\
\cmidrule(lr){2-9}
& \multirow{5}{*}{Thomas}
& MLE           & -0.027 (0.09) & \textbf{-0.007 (0.03)}
                & \textbf{-0.027 (0.11)} & \textbf{-0.006 (0.03)}
                & -0.064 (0.04) & -0.011 (0.02) \\
& & INLA-SLP    & NA* & -0.015 (0.03)
                & 237.511 (5311.04) & -0.016 (0.03)
                & \textbf{-0.005 (0.15)} & \textbf{0.003 (0.03)} \\
& & ISIW Known  & 0.026 (0.19) & -0.012 (0.04)
                & 0.066 (0.25) & -0.010 (0.04)
                & 0.010 (0.14) & -0.013 (0.04) \\
& & ISIW KIE    & \textbf{0.018 (0.18)} & -0.010 (0.05)
                & 0.053 (0.24) & -0.008 (0.06)
                & -0.054 (0.06) & -0.012 (0.03) \\
& & ISIW KIE COV& 0.066 (0.20) & -0.009 (0.04)
                & 0.091 (0.23) & -0.008 (0.05)
                & 0.027 (0.14) & -0.018 (0.05) \\
\midrule
%%%%%%%%%%%%%%%%%%%%%%%%%%%%%%%%%%%%%%%%%%%%%%%%%%%%%%%%%%%%%%%%%%%%%%%%%%%%
%%%%%%%%%%%%%%%%%%%%%%%%%%%%%%% RMSE %%%%%%%%%%%%%%%%%%%%%%%%%%%%%%%%%%%%%%%
%%%%%%%%%%%%%%%%%%%%%%%%%%%%%%%%%%%%%%%%%%%%%%%%%%%%%%%%%%%%%%%%%%%%%%%%%%%%
\multirow{10}{*}{RMSE}
& \multirow{5}{*}{LGCP}
& MLE           & \textbf{0.108 (0.10)} & \textbf{0.027 (0.02)}
                & \textbf{0.101 (0.08)} & \textbf{0.026 (0.02)}
                & \textbf{0.056 (0.04)} & 0.022 (0.02) \\
& & INLA-SLP    & 0.127 (0.47) & 7.471 (166.39)
                & NA* & 0.030 (0.02)
                & 0.089 (0.50) & \textbf{0.020 (0.02)} \\
& & ISIW Known  & 0.167 (0.19) & 0.039 (0.03)
                & 0.185 (0.25) & 0.040 (0.03)
                & 0.138 (0.14) & 0.042 (0.03) \\
& & ISIW KIE    & 0.190 (0.22) & 0.050 (0.05)
                & 0.214 (0.26) & 0.059 (0.07)
                & 0.144 (0.18) & 0.039 (0.04) \\
& & ISIW KIE COV& 0.151 (0.15) & 0.035 (0.03)
                & 0.183 (0.19) & 0.035 (0.03)
                & 0.149 (0.12) & 0.038 (0.03) \\
\cmidrule(lr){2-9}
& \multirow{5}{*}{Thomas}
& MLE           & \textbf{0.082 (0.05)} & \textbf{0.023 (0.02)}
                & \textbf{0.087 (0.07)} & \textbf{0.023 (0.02)}
                & \textbf{0.069 (0.03)} & \textbf{0.021 (0.02)} \\
& & INLA-SLP    & NA* & 0.030 (0.02)
                & 237.640 (5311.03) & 0.028 (0.02)
                & 0.073 (0.13) & 0.025 (0.02) \\
& & ISIW Known  & 0.124 (0.14) & 0.032 (0.02)
                & 0.163 (0.21) & 0.033 (0.03)
                & 0.094 (0.10) & 0.035 (0.03) \\
& & ISIW KIE    & 0.115 (0.14) & 0.038 (0.04)
                & 0.145 (0.20) & 0.040 (0.04)
                & 0.069 (0.04) & 0.027 (0.02) \\
& & ISIW KIE COV& 0.150 (0.14) & 0.036 (0.03)
                & 0.173 (0.18) & 0.035 (0.03)
                & 0.111 (0.09) & 0.043 (0.03) \\
\bottomrule
\end{tabular}}
\end{table}

\begin{table}[h]
\centering
\caption{Mean RMSPE (SE) across the $B=500$ simulations for the point process misspecification simulation experiment with $N=100$ points. Bold indicates the best (lowest) value within each point process (LGCP or Thomas) for each simulation scenario.}
\label{tab:replication}
\resizebox{\textwidth}{!}{
\begin{tabular}{ll cccc cc}
\toprule
\textbf{Point process}& \textbf{Method} & \multicolumn{2}{c}{$\boldsymbol{\beta} = -1$} & \multicolumn{2}{c}{$\boldsymbol{\beta} = 1$} & \multicolumn{2}{c}{$\boldsymbol{\beta} = 2$} \\
\cmidrule(lr){3-4} \cmidrule(lr){5-6} \cmidrule(lr){7-8}
& & $\phi = 0.02$ & $\phi = 0.15$ & $\phi = 0.02$ & $\phi = 0.15$ & $\phi = 0.02$ & $\phi = 0.15$ \\
\midrule
\multirow{5}{*}{LGCP}
& MLE & 1.731 (0.11) & 0.780 (0.18) & 1.725 (0.11) & 0.782 (0.17) & 2.533 (0.15) & 1.387 (0.36) \\
& INLA-SLP & 1.212 (0.12) & \textbf{0.552 (0.08)} & 1.206 (0.09) & \textbf{0.553 (0.08)} & \textbf{1.417 (0.16)} & \textbf{0.701 (0.15)} \\
& ISIW Known & \textbf{1.190 (0.10)} & 0.660 (0.14) & \textbf{1.190 (0.11)} & 0.660 (0.13) & 1.598 (0.36) & 1.014 (0.33) \\
& ISIW KIE & 1.596 (0.14) & 0.718 (0.16) & 1.594 (0.15) & 0.733 (0.21) & 2.325 (0.25) & 1.235 (0.37) \\
& ISIW KIE COV & 1.697 (0.12) & 0.734 (0.16) & 1.697 (0.12) & 0.734 (0.15) & 2.550 (0.25) & 1.260 (0.36) \\

\midrule
\multirow{5}{*}{Thomas}
& MLE & 1.664 (0.12) & 0.820 (0.20) & 1.661 (0.11) & 0.820 (0.20) & 2.219 (0.25) & 1.374 (0.46) \\
& INLA-SLP & 1.377 (0.36) & 0.770 (0.29) & 1.292 (0.17) & 0.749 (0.24) & 2.897 (1.56) & 1.281 (0.78) \\
& ISIW Known & \textbf{1.197 (0.08)} & \textbf{0.710 (0.15)} & \textbf{1.204 (0.10)} & \textbf{0.709 (0.15)} & \textbf{1.462 (0.31)} & \textbf{1.078 (0.38)} \\
& ISIW KIE & 1.540 (0.16) & 0.766 (0.18) & 1.542 (0.15) & 0.763 (0.18) & 2.101 (0.29) & 1.256 (0.47) \\
& ISIW KIE COV & 1.637 (0.15) & 0.773 (0.18) & 1.648 (0.10) & 0.769 (0.18) & 2.234 (0.33) & 1.282 (0.53) \\
\bottomrule
\end{tabular}}
\end{table}

\clearpage

% \begin{figure}[t]
% \centering
% \includegraphics[scale=0.96]{figures/aim2-ISIW/ISIW-Point-Processes.pdf}
% \caption{One realization of the point processes used in the simulation analysis for $n=200$. The top and bottom row fields are low range ($\phi=0.02$) and high range ($\phi=0.15$) respectively.}
% \label{fig:aim2_point_proces}
% \end{figure}

\section{Additional real data analysis}

\subsection{Galicia moss data}

\begin{figure}[h]
\centering
\includegraphics[scale=1]{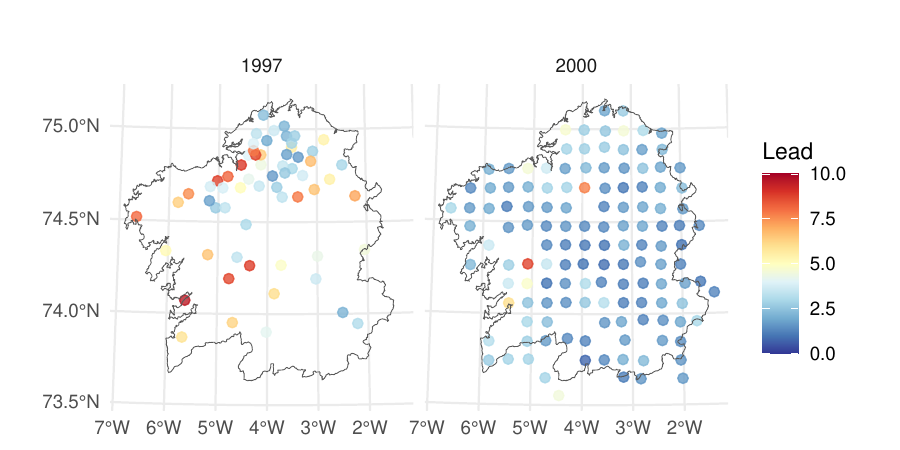}
\caption{Observed lead concentrations at sampled locations in Galicia in 1997 and 2000.}
\label{fig:aim2_galicia_data}
\end{figure}

% Table: Galicia model results
\begin{table}[h]
    \centering
    \caption{Parameter estimates for the Galicia data.}
    \label{tab:galicia_param}
    \renewcommand{\arraystretch}{1.2}
    \begin{tabular}{llrrrrr}
        \toprule
        Dataset & Method & $\mu$ & $\sigma^2$ & $\phi$ & $\tau^2$ & $\beta$ \\
        \midrule
        \multirow{4}{*}{Galicia}
        & MLE 
            & 1.546 
            & 0.123 
            & 0.142 
            & 0.108 
            & -- \\
        & ISIW KIE 
            & 1.958 
            & 0.0418 
            & 0.785 
            & 0.0619 
            & -- \\
        & ISIW KIE COV 
            & 1.581 
            & 0.0468 
            & 0.189 
            & 0.0854 
            & -- \\
        & INLA-SLP 
            & 2.164 
            & 0.137 
            & 0.863 
            & 0.192 
            & -5.43 \\
        \bottomrule
    \end{tabular}
\end{table}

\clearpage

\subsection{California AQS data}
\begin{figure}[h]
\centering
\includegraphics[width=\textwidth]{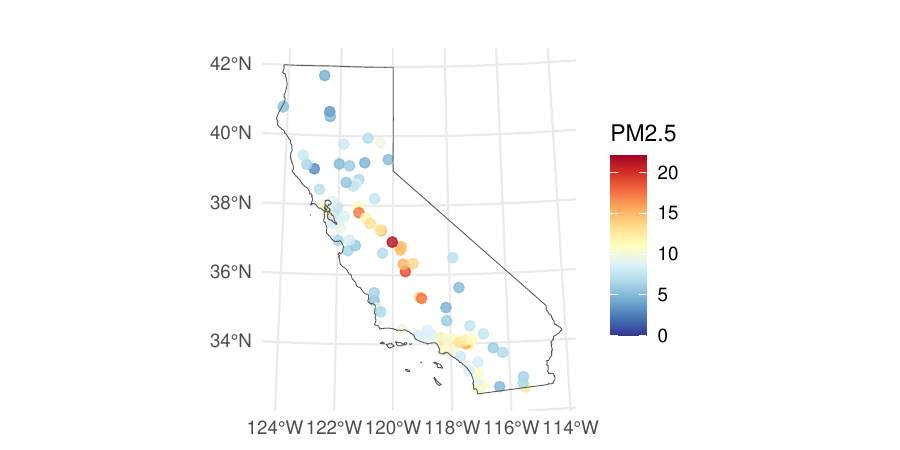}
\caption{Average daily concentrations of $\text{PM}_{2.5}$ at monitor locations in California in 2010.}
\label{fig:ca_aqs_data}
\end{figure}

% Table: Galicia model results
\begin{table}[h]
    \centering
    \caption{Parameter estimates for the California AQS data.}
    \label{tab:ca_aqs_param}
    \renewcommand{\arraystretch}{1.2}
    \begin{tabular}{llrrrrr}
        \toprule
        Dataset & Method & $\mu$ & $\sigma^2$ & $\phi$ & $\tau^2$ & $\beta$ \\
        \midrule
        \multirow{4}{*}{California AQS}
        & MLE 
            & 2.020 
            & 0.134 
            & 0.472 
            & 0.0146 
            & -- \\
        & ISIW KIE 
            & 1.805 
            & 0.102 
            & 0.528 
            & 0.0275 
            & -- \\
        & ISIW KIE COV 
            & 1.944 
            & 0.0597 
            & 0.330 
            & 0.0276 
            & -- \\
        & INLA-SLP 
            & 0.831 
            & 0.643 
            & 1.260 
            & 0.0167 
            & 2.43 \\
        \bottomrule
    \end{tabular}
\end{table}

% Figure: Galicia model results
% \begin{table}[h]
%     \centering
%     \caption{Parameter estimates for the log-transformed Galicia data.}
%     \label{tab:galicia_param}
%     \renewcommand{\arraystretch}{1.2}
%     \begin{tabular}{llrrrrr}
%         \toprule
%         Year & Method & $\mu$ & $\sigma^2$ & $\phi$ & $\tau^2$ & $\beta$\\
%         \midrule
%         \multirow{6}{*}{1997} 
%         & MLE      & 1.55  & 0.123  & 0.142  & 0.108 & - \\
%         & ISIW-V   & 1.93  & 0.032  & 0.678   & 0.066 & - \\
%         & ISIW-PM  & 1.73  & 0.098  & 0.050   & 0.003 & - \\
%         & INLA-SLP [$P(\sigma^2>1)=0.01]$ & 2.41 & 0.385 & 2.23  & 0.198  & -4.38\\
%         & INLA-SLP [$P(\sigma^2>0.5)=0.01$] & 2.89 & 1.09 & 0.194  & 0.000 & -1.60 \\
%         & INLA-SLP [$P(\sigma^2>0.1)=0.01$] & 0.956 & 0.079 & 3.13  & 0.247 & 7.85 \\
%         \midrule
%         \multirow{6}{*}{2000} 
%         & MLE      & 0.716  & 0.130  & 0.185   & 0.037 & -\\
%         & ISIW-V   & 0.756  & 0.137  & 0.167   & 0.031  & -\\
%         & ISIW-PM  & 0.689  & 0.144  & 0.145   & 0.021  & -\\
%         & INLA-SLP [$P(\sigma^2>1)=0.01$] & 3.43  & 2.54  & 2.95   & 0.080 & -1.05 \\
%         & INLA-SLP [$P(\sigma^2>0.5)=0.01$] & 3.06 & 1.70 & 2.61  & 0.082 & -1.17 \\
%         & INLA-SLP [$P(\sigma^2>0.1)=0.01$] & -1.21 & 0.937 & 3.41  & 0.098 & 1.21 \\
%         \bottomrule
%     \end{tabular}
% \end{table}

\clearpage

\subsection{Prior sensitivity analysis for INLA-SLP}

We conducted a small prior sensitivity analysis for the Galicia data to illustrate how spatial predictions are very sensitive to the prior on the variance parameter. Reference values of 0.1, 0.5, and 1 were examined in the PC prior, selected based on the Empirical Bayes estimate obtained from the INLA-SLP fit to the Galicia data ($\hat{\sigma}^2 = 0.146$). These values were chosen to span both smaller and larger ranges relative to the empirical estimate, thereby assessing the sensitivity of spatial predictions to the prior specification.

\begin{figure}[h]
\centering
\includegraphics[scale=1]{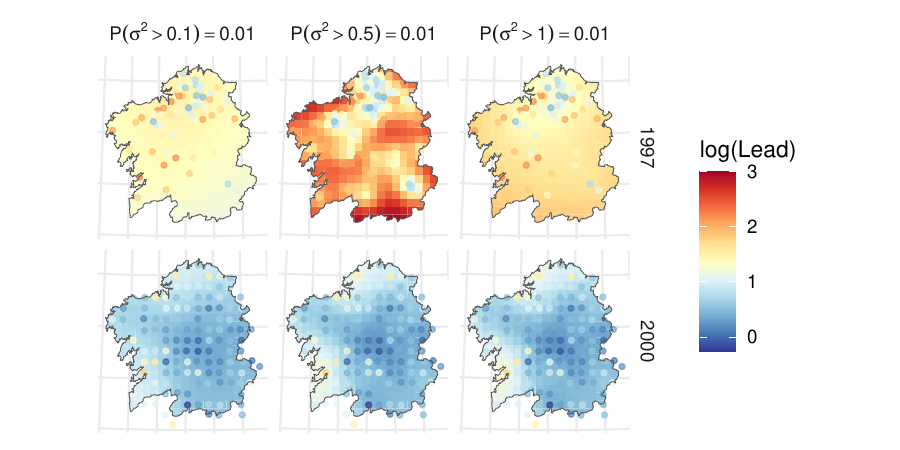}
\caption{Spatial predictions of log lead concentrations in Galicia in 1997 and 2000 using INLA-SLP for three different PC priors on the variance parameter ($\sigma^2$). Points represent the observed data.}
\label{fig:aim2_galicia_pred_inla}
\end{figure}

\end{document}